\pdfoutput=1
\documentclass[final]{elsart3p}
\usepackage{graphicx}
\usepackage[center]{subfigure}

\newcommand{\deriv}[2]{\ensuremath{\frac{\partial #1}{\partial #2}}}
\newcommand{\code}[1]{\texttt{#1}}

\newcommand{\pow}{\ensuremath{\wedge}}

\newcommand{\apar}{\ensuremath{A_{||}}}
\newcommand{\hthe}{\ensuremath{h_\theta}}
\newcommand{\Bp}{\ensuremath{B_\theta}}
\newcommand{\Bt}{\ensuremath{B_\zeta}}
\newcommand{\Vec}[1]{\ensuremath{\mathbf{#1}}}
\newcommand{\bvec}{\Vec{b}}
\newcommand{\kvec}{\Vec{\kappa}}
\newcommand{\bxk}{\bvec_0\times\kvec_0\cdot\nabla}

\newcommand{\Jpar}{J_{||}}
\newcommand{\delp}{\nabla_\perp^2}

\newcommand{\Curl}[1]{\ensuremath{\nabla\times #1 }}

\journal{Computer Physics Communications}

\begin{document}
\begin{frontmatter}

\title{BOUT++: a framework for parallel plasma fluid simulations}
\author{B.D.Dudson\corauthref{cor}}
\corauth[cor]{Corresponding author.}
\ead{bd512@york.ac.uk}
\author{, H.R.Wilson}
\address{Department of Physics, University of York, Heslington, York YO10 5DD, UK}

\author{M.V.Umansky}
\author{, X.Q.Xu}
\address{Lawrence Livermore National Laboratory, Livermore, CA 94551, USA}

\author{P.B.Snyder}
\address{General Atomics, P.O. Box 85608, San Diego, CA 92186-5608, USA}

\begin{abstract}
A new modular code called BOUT++ is presented, which simulates 3D fluid equations in curvilinear coordinates.
Although aimed at simulating Edge Localised Modes (ELMs) in tokamak x-point geometry, the
code is able to simulate a wide range of fluid models (magnetised and unmagnetised)
involving an arbitrary number of scalar and vector fields, in a wide range of geometries. 
Time evolution is fully implicit, and $3^{rd}$-order WENO schemes are implemented.
Benchmarks are presented for linear and non-linear
problems (the Orszag-Tang vortex) showing good agreement. Performance of the
code is tested by scaling with problem size and processor number, showing efficient scaling
to thousands of processors. 

Linear initial-value simulations of ELMs using reduced ideal MHD are presented,
and the results compared to the ELITE linear MHD eigenvalue code. 
The resulting mode-structures and growth-rate are found to be in good agreement
($\gamma_{BOUT++}=0.245\omega_A$, $\gamma_{ELITE} = 0.239\omega_A$, with  Alfv\'enic timescale $1/\omega_A = R/V_A$). 
To our knowledge, this is the first time dissipationless, initial-value simulations of ELMs have been 
successfully demonstrated.
\end{abstract}


\end{frontmatter}

\section{Introduction}
BOUT++ is a new highly adaptable, object-oriented C++ code for performing parallel plasma
fluid simulations with an arbitrary number of equations in 3D curvilinear coordinates
using finite-difference methods. 
It has been developed from the original {\bf BOU}ndary {\bf T}urbulence 3D 2-fluid
tokamak edge simulation code BOUT \cite{bout_manual,xu-1998,xu-1999,xu-2000-pop,rognlien-2002,xu-2008},
borrowing ideas and algorithms, but has been substantially altered and extended.
Though designed to simulate tokamak edge plasmas efficiently, the methods used are very general and
can be adapted to many other situations: any coordinate system metric tensor 
$g^{ij} = g^{ij}\left(x,y\right)$ (i.e. constant in one dimension) can be specified, which
restricts the coordinate system to those with axi- or translationally symmetric geometries.
Even 2D metric tensors encompass a wide range of situations, allowing the code to be
used to simulate plasmas in slab, sheared slab, cylindrical
and non-orthogonal coordinate systems such as flux coordinates for tokamak simulations. 
Extension of the code to allow 3D metric tensors would be relatively straightforward,
but is not currently necessary for the problems considered here.

BOUT++ is designed to automate the common tasks needed for fluid finite-difference
simulation codes, and to separate the complicated (and error-prone) details such as
differential geometry, parallel communication, and file input/output from the
user-specified physics equations to be solved, whilst remaining as flexible as possible. 
Thus the physics equations being solved
are clearly provided in one place, and can be easily changed with only minimal knowledge of the inner workings of the
code. As far as possible, this allows the user to concentrate on the physics, rather than worrying
about the numerics.

\subsection{Related work}

Several frameworks for parallel simulation already exist, for example POOMA and Overture
(both parts of the Advanced CompuTational Software (ACTS) collection \cite{acts}). These are
very flexible and powerful, but still require significant knowledge of parallel computing,
and investment of time, to produce a working simulation. BOUT++ provides a more complete framework,
significantly reducing the time needed and allowing quick testing of methods and equations.

Very similar in spirit to BOUT++ is the OpenFOAM project \cite{weller-1998,jasak-1998}.
This is an unstructured mesh finite-volume code, which also uses C++ features such as operator overloading
to simplify the process of creating new simulations. The main distinction is that whereas
OpenFOAM is designed to simulate complex shaped domains in Cartesian coordinates, BOUT++
simulates relatively topologically simple domains in complicated coordinate systems. Therefore,
OpenFOAM is more suited to engineering applications such as simulation
of internal combustion engines \cite{jasak-2004}, whilst BOUT++ is more suited to problems in
physics such as simulation of tokamaks where non-Cartesian coordinate systems can be used to
exploit anisotropies and so optimise numerical solution.

In this paper the present state of the BOUT++ code is described, with the general layout of the code
discussed in section~\ref{sec:layout}, after a brief introduction to the physics motivation in
section~\ref{sec:phys_overview}.
Details of the numerical methods implemented are described
in section~\ref{sec:difops}. A series of test problems are used to demonstrate
the accuracy and flexibility of the code in section~\ref{sec:tests}.
Since the most immediate application of BOUT++ is to problems in plasma physics,
the test problems are drawn from this field.
Simulation of Edge Localised Modes (ELMs) and comparison with the ELITE linear MHD eigenvalue code
\cite{wilson-2002,snyder-2002} are presented in section~\ref{sec:elmsim}.
Finally, the performance of the BOUT++ code is demonstrated in section~\ref{sec:perform},
showing efficient scaling with number of processors for a fixed problem size
(hard scaling) to thousands of processors in section~\ref{sec:scale_proc}.

\section{Physics overview}
\label{sec:phys_overview}

\begin{figure}[htb!]
\centering
\includegraphics[scale=0.5]{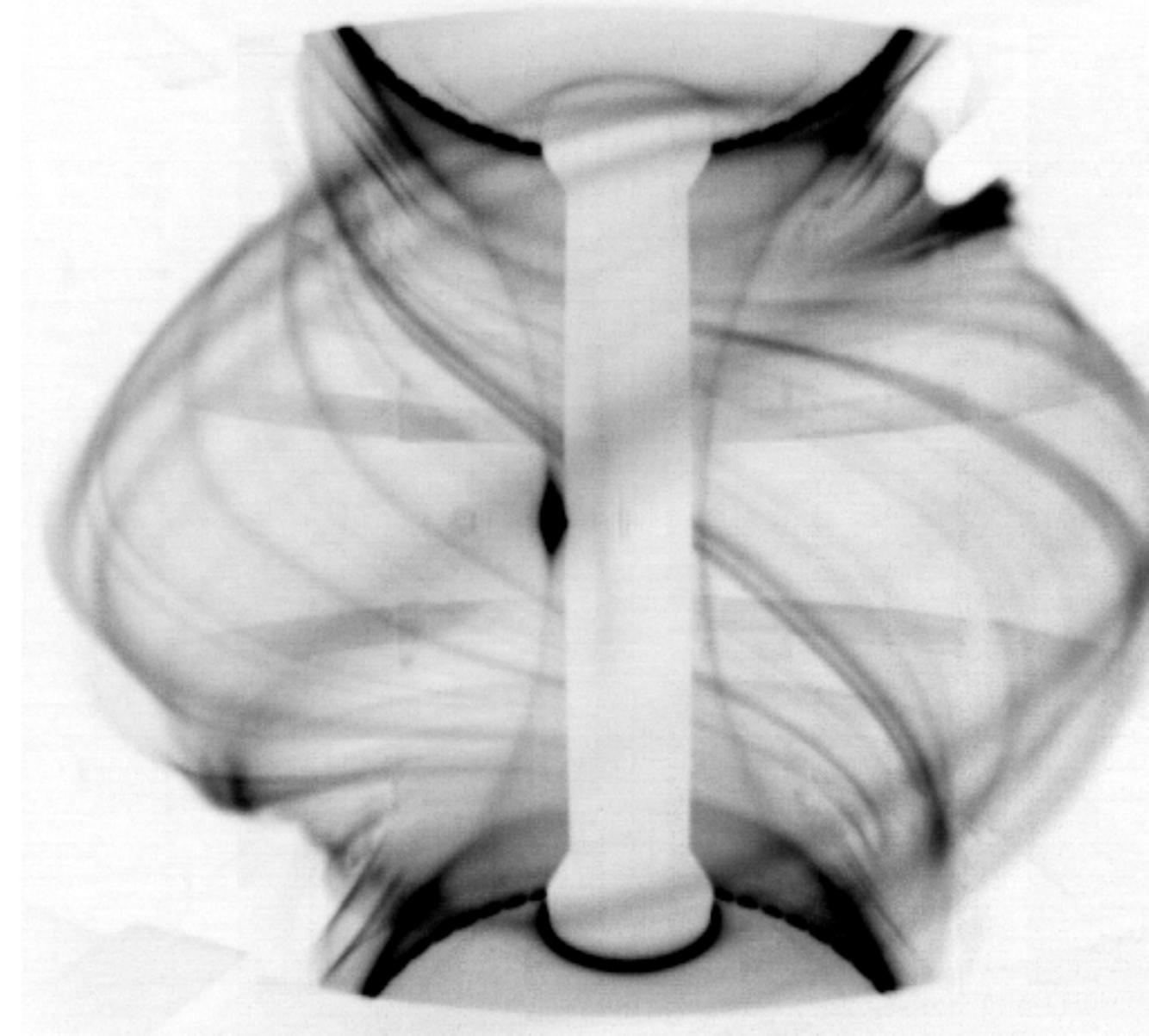}
\caption{Negative image of an Edge Localised Mode in the MAST Tokamak \cite{kirk-2006}, showing eruption of filaments from the plasma edge}
\label{fig:elm_image}
\end{figure}

Edge Localised Modes in tokamaks are sudden (sub-millisecond) releases
of particles and energy, resulting in the eruption of filamentary structures from the plasma edge.
An image of one of these events from the Mega-Amp Spherical Tokamak \cite{kirk-2006} using a $10~\mu s$ exposure
time is shown in figure~\ref{fig:elm_image}.
The particles and energy released during these events are deposited on material surfaces and
are potentially damaging in future fusion devices. There is therefore interest in understanding and
controlling these events.

ELMs are found to be triggered close to the stability boundary of an ideal magnetohydrodynamic (MHD) mode,
called the peeling-ballooning mode \cite{hegna-1996,snyder-2002}. This provides strong evidence that this mode
is involved in triggering an ELM. Peeling-ballooning modes are destabilised by a combination of pressure
gradients (ballooning) and currents close
to the plasma edge (peeling) \cite{connor-1998}. Further details of the
linear structure of peeling-ballooning modes are given in section~\ref{sec:elmsim}
where BOUT++ simulations of this mode are discussed. 
Although there are analytic theories \cite{wilson-2004}
and semi-analytic models \cite{zhu-2007} for the early non-linear evolution of this mode,
it is not yet fully understood how this develops into the filamentary structures observed,
and ultimately how particles and energy are lost.

Several 3D non-linear codes have been used to simulate ELMs, including NIMROD \cite{sovinec-2004,brennan-2006,pankin-2007},
BOUT \cite{snyder-2005,snyder-2006}, JOREK \cite{huysmans-2007}, GEM \cite{scott-2005,scott-2006} and M3D \cite{park-1999,park-2007}. These codes incorporate a wide range of physics
including (in the case of BOUT and some NIMROD simulations e.g. \cite{sovinec-2007}) 2-fluid effects.
The approach employed with these codes
is essentially to reproduce experimental observations and then to relate these results back to analytic theory.
A complementary approach is to attack the problem from the other direction: starting from the analytic theory
(i.e. ideal MHD models), gradually build complexity into the model in order to approach experimental results.
The BOUT++ code is being developed and benchmarked to follow this second approach. For this reason the BOUT++
code has been designed to be very flexible in order to allow rapid prototyping of simulations involving different
physical models. This is useful because it is not yet known what physical effects are essential
to adequately simulate an ELM, or what numerical methods are most suited to the problem.

\section{Structure of the program}
\label{sec:layout}

The BOUT++ code can be separated into the following components:
\begin{itemize}
\item Time integration using the Sundials CVODE package \cite{acts} (section~\ref{sec:time_int}).
\item Input and output to the Portable Data Binary (PDB) format \cite{pact-llnl} (section~\ref{sec:io}).
\item Low-level data handling: encapsulation of variables into objects
  with associated operators (section~\ref{sec:data}).
\item Parallel communications using the Message Passing Interface (MPI)
  (section~\ref{sec:comms})
\item Coordinate system and differential operators (section~\ref{sec:coords})
\item Differencing methods, both central and upwinding (section~\ref{sec:difops})
\item The physics module which determines the equations to be solved (section~\ref{sec:physics_model})
\end{itemize}
Each of these components can be modified without altering the other modules, provided that
the interface methods are the same. In particular, the physics module has been designed
to be the easiest to replace since this is the one most users will need to alter.
We now detail each of these components, and the algorithms used.

\subsection{Time integration}
\label{sec:time_int}

\begin{figure}[htb!]
\centering
\includegraphics[scale=0.5]{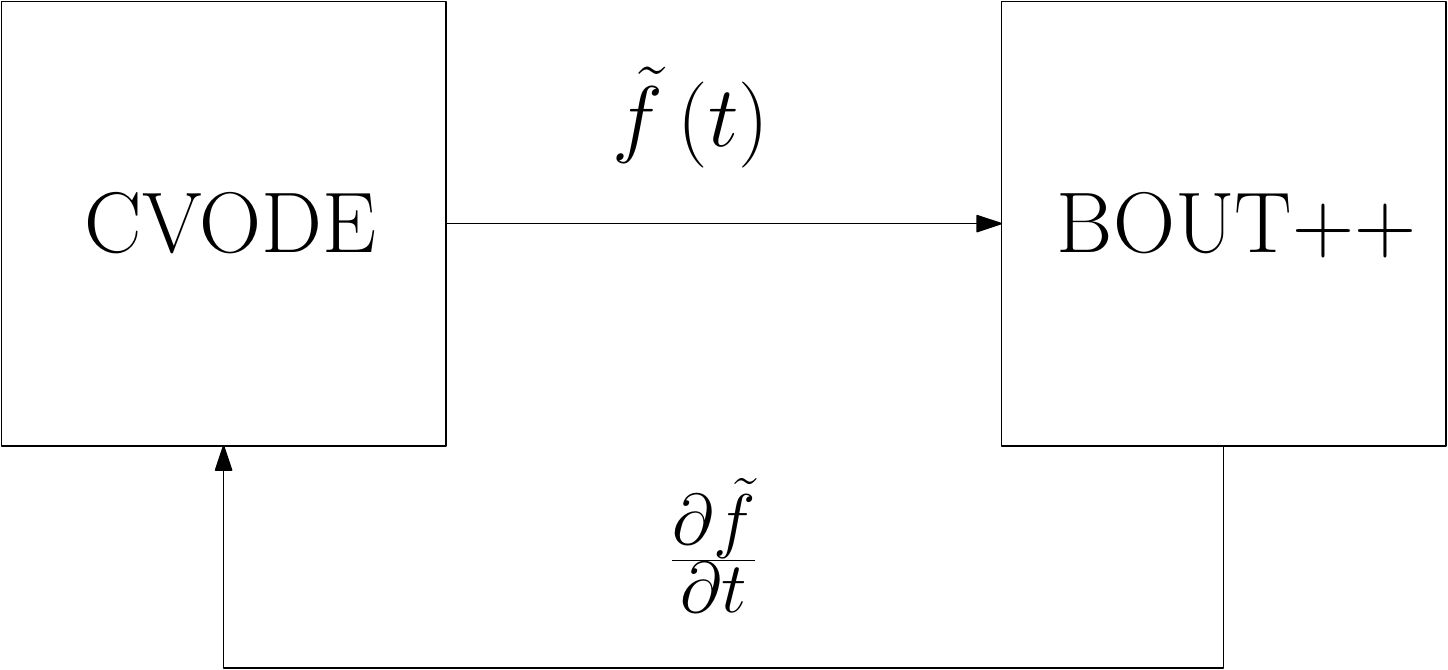}
\caption{Data flow between CVODE library and BOUT++ code}
\label{fig:pvode}
\end{figure}

Like BOUT, BOUT++ is built upon the general implicit time-integration solver CVODE \cite{acts}. 
This is used as a ``black-box'', requiring no information about the equations themselves,
only the values of the time-derivatives given a state of the system.
This is illustrated in figure~\ref{fig:pvode}: the state of the system at a given time
$\tilde{f}\left(t\right)$ is passed from the CVODE library to BOUT++. From this, BOUT++
calculates the time-derivatives of all quantities $\partial \tilde{f} / \partial t$
which is passed back to CVODE.
This independence of the equations to be solved makes CVODE an ideal starting point 
for creating a general simulation framework.

To advance the system state in time, CVODE uses the Newton-Krylov BDF implicit method
for stiff problems. To be efficient, this method requires preconditioning and this can
optionally be supplied to the solver. It has been found however that for the simulations
so-far attempted, this has not been necessary.
The incorporation and exploration of physics-based preconditioners is planned as a
future extension.

The interface to CVODE is in C, and so this has been wrapped into a C++ class. In principle
therefore the solver could be replaced by a different package without affecting the remaining
code.

\subsection{Input and output}
\label{sec:io}

Input to BOUT++ consists of two files: an options text file, and a binary grid file.
The options file format is the same as a windows INI file, consisting of a mixture
of comments, section headers and \code{name = value} pairs which makes the settings used for 
a given simulation clear. All aspects of a simulation can be set at run-time except the
equations solved which are set in a compiled physics module (section~\ref{sec:physics_model}).
This includes the number of steps, run-time limits, data and restart output period, differencing methods,
field initialisation and boundary conditions. Use of compile-time options
(\#define C preprocessor directives)
tends to confuse which settings were used for a given simulation and so these are not used
except for debugging options (section~\ref{sec:debug}). Instead, by keeping all
options in one file and assigning default values to new options, 
simulations can be easily repeated at a later time even if the code has changed internally.

Binary data input and output (grid input, data and restart file output) are in the
Portable Data Binary (PDB) format \cite{pact-llnl}. For pre- and post-processing of
input and output files C, FORTRAN and python bindings are supplied as part of the Portable Application Code Toolkit
\cite{pact-llnl}, and an IDL library has been separately developed. 
IDL and python bindings in particular, enable fast development
of interactive codes to view and analyse results.
Future development includes the possibility of moving to a more widely used binary
format such as netCDF \cite{brown-1993,netcdf-www} or HDF5 \cite{hdf5-www}. 

\subsection{Data handling}
\label{sec:data}

The simplest part of a simulation code is the handling of data storage and manipulation,
but is also time-consuming and error-prone. In BOUT++ this is handled by a set of classes
which manage all memory management and looping over domain indices, allowing the
remainder of the code to be written in a much more concise manner.
Operator overloading allows 3D scalar and vector fields to be treated as simple variables,
eliminating some common bugs due to mis-typing
array indices, and making the source code much easier to read.

Several data classes have been
implemented: 3D scalar and vector fields, and axisymmetric (2D) scalar and vector fields which are
constant in the z coordinate and are useful for tokamak simulations because all equilibrium quantities
are axisymmetric in toroidal angle (see section~\ref{sec:tok_coord}). Scalar overloaded
operations include addition, multiplication, exponentiation by real values or scalar fields. For vector
fields, the dot and cross products are represented by \code{*} and \pow symbols.
For example, the following examples are all valid operations on scalar fields \code{a,b} and \code{c},
and vector fields \code{x} and \code{y}:
\begin{verbatim}
Scalar3D a, b, c;  // 3D scalar fields
real r;

a = b + c; a = b - c;             // Addition & Subtraction
a = b * c; a = r * b;             // Multiplication
a = b / c; a = b / r; a = r / b;  // Division
a = b ^ c; a = b ^ r; a = r ^ b;  // Exponentiation

Vector3D x, y, z;  // 3D vector fields

x = y * a; // Multiplication by scalar field

a = x * y // Dot-product
x = y ^ z // Cross-product
\end{verbatim}

For both scalar and vector field operations, so long as the result of an operation is of the correct type,
the usual C/C++ shorthand notation can be used (i.e. \code{a *= b} is equivalent to \code{a = a * b}).
These operations can of course be combined into statements such as \code{a = 4*b + (c\pow 2)}.
A complication is that in C++ the \pow  ~operator has lower precedence than the \code{*} or \code{+} operators
and so exponentiation and cross-product operations must be put in brackets.

In addition to arithmetic operations, standard mathematical functions such as \code{sqrt()} and \code{abs()}
are also overloaded. This allows all operations on scalar and vector fields to be written very clearly and
concisely.

\subsubsection{Optimisation}
\label{sec:opt}

In most cases, a hand-coded, specialised program will execute faster than a more flexible code.
Since flexibility is an aim of BOUT++, and performance is a concern for large-scale simulations, 
this must be addressed. A famous quote by Donald Knuth goes ``We should forget about
small efficiencies, say about 97\% of the time: premature optimisation is the root of all evil.'' \cite{knuth-1974}, 
i.e. performance should not be the guiding principle in designing a code. This is  because optimisations
treat special cases, making code less clear and bugs harder to find. Whilst developing BOUT++
it has been generally found that high-level algorithms have a greater effect on execution times
than low-level operations. In this case a small performance penalty is worthwhile because the flexibility
gained allows faster development of high-level algorithms.

In optimising BOUT++, bottlenecks have been identified using profiling tools, and 
optimisations made where these did not adversely affect the clarity of the code. 
Two optimisations used in the data objects to speed up code execution are memory recycling, which eliminates
allocation and freeing of memory; and copy-on-change, which minimises unnecessary copying of data.

Both of these optimisations are done ``behind the scenes'', hidden from the remainder of the code, and
are illustrated in figure~\ref{fig:memory}:
\begin{figure}[htb!]
\centering
\includegraphics[scale=0.4]{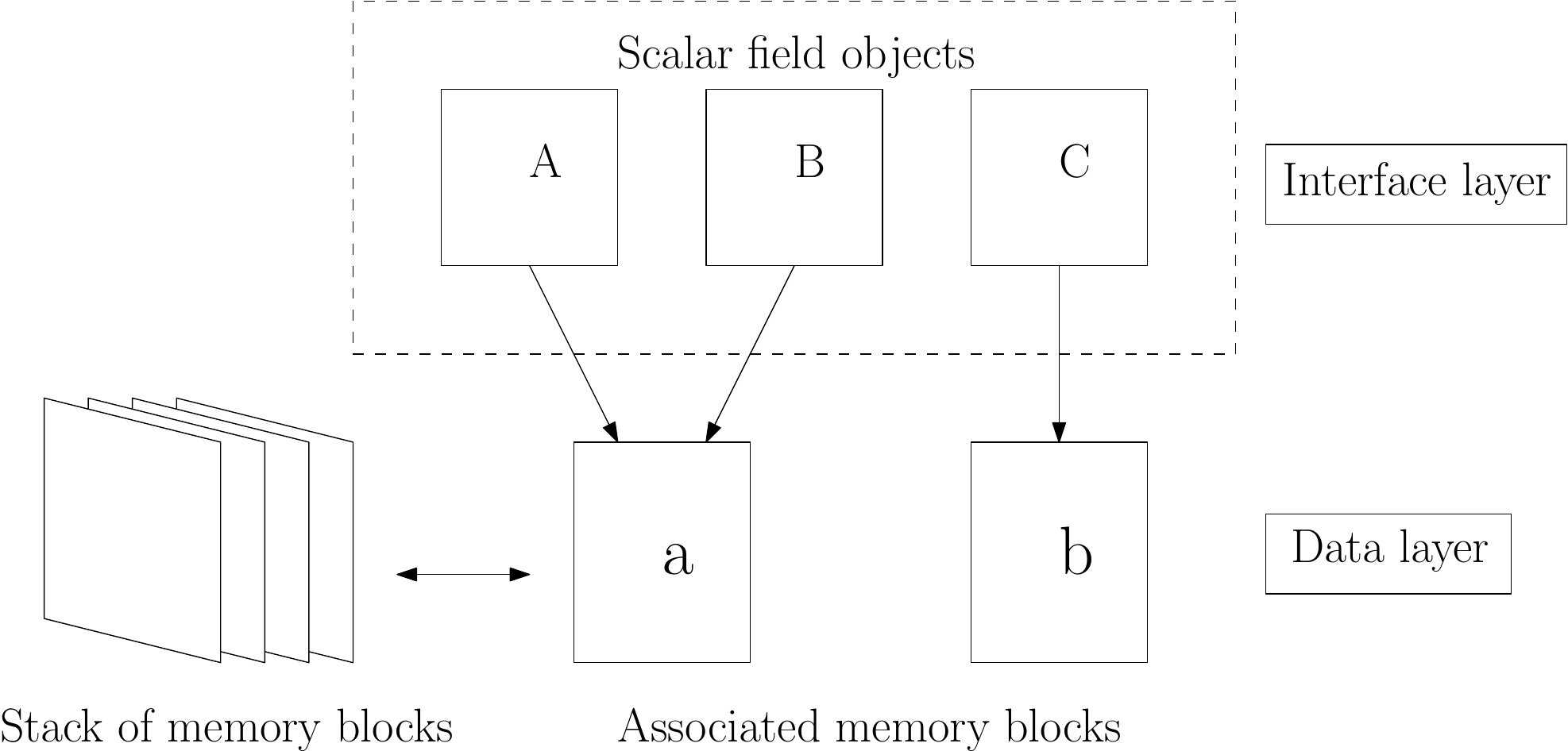}
\caption{Memory handling in BOUT++}
\label{fig:memory}
\end{figure}
The objects (A,B,C) accessed by the user in operations
discussed in the previous section act as an interface to underlying data (a,b). 
Memory recycling can be used because all the scalar fields are the same size (and vector fields are
implemented as a set of 3 scalar fields). Each class implements a global stack of available
memory blocks. When an object is assigned a value, it attempts to grab one of these memory blocks,
and if none are available then a new block is allocated. 
When an object is destroyed, its memory block is not freed, but is put onto the
stack. Since the evaluation of the time-derivatives involves the same set of operations each time, this system
means that memory is only allocated the first time the time-derivatives are calculated, after which the same 
memory blocks are re-used. This eliminates the often slow system calls needed to allocate and free memory,
replacing them with fast pointer manipulation. 

Copy-on-change (reference counting) further reduces memory usage and unnecessary copying of data. 
When one field is set equal to another (e.g. \code{Scalar3D A = B} in figure~\ref{fig:memory}), no 
data is copied, only the reference to the underlying data (in this case both A and B point to data block a). 
Only when one of these objects is modified is a second memory block used to store the different value. 
This is particularly useful when returning objects from a routine. Usually this would
involve copying data from one object to another, and then destroying the original copy. Using
reference counting this copying is eliminated.

\subsubsection{Debugging support}
\label{sec:debug}

Several features are built into the BOUT++ data objects which aid debugging,
and can be enabled at compile-time: run-time checking, operation and variable
tracking, and segmentation fault handling. Run-time checking tests the result
of every operation (or subsets, depending on checking level)
for non-finite results (\code{NaN, Inf}), stopping with an error message when
such a value is encountered. In order to help locate where these values have
occurred, an additional compile flag can be used to enable tracking of
operations: variables can be assigned names, and the result of an operation on
two fields is given a name based on the input names. For example, the result
of \code{A+B} would be given the name ``(A+B)'', and similarly for all other
operations and differential operators. Thus, when an error occurs the name of
the variables involved can be printed; for example an error might read
``\code{Scalar3D: Non-finite number at [12][2][10] in `sqrt(a - b)'}''.
These options can be used for initial testing of a module, and then disabled
for long production simulations.

Tracking down bugs in a large code like BOUT++ can be very tricky, particularly for
intermittent problems such as segmentation faults. This is because
these can be impractical to reproduce running under a debugger due to the run-time, and may even
not occur under a debugger due to the different memory layout. Finding where a bug
occurs can therefore take a lot of trial-and-error. To help catch this kind of bug, a fast message stack has been
implemented in BOUT++: the idea is that at the start of every function (or part of a function) which
may cause faults a message is put in a stack, and then removed once the function finishes. If
an error is found - either from non-finite number checking or segmentation fault - the message
stack is printed out, giving a good idea of where the error occurred. The following is an example
of a segmentation fault deliberately triggered in a parallel derivative operation in the RHS function:

\begin{verbatim}

****** SEGMENTATION FAULT CAUGHT ******

====== Back trace ======
 -> Grad_par( Scalar3D )
 -> Running RHS: Solver::rhs(0.000000e+00)
 -> Initialising solver

\end{verbatim}

The penalty for enabling run-time checking for most operations, and the message stack above is an
increase in run-time of 10-15\%. For most simulations so far performed, this computational cost
has been acceptable, and so checking was enabled for all calculations presented here, including
scaling runs in section~\ref{sec:perform}. For longer simulations once a code has been
thoroughly tested, the cost of run-time checking may become problematic. For these runs, 
all checking can be disabled.

\subsection{Communication and topology}
\label{sec:comms}

Though parallel communication could be handled transparently by the data objects, there are several potential
efficiency improvements which would be difficult to automate, such as overlapping communication with calculations.
For this reason, parallel communications in BOUT++ are handled by a separate object to provide this flexibility.
Field objects are grouped into communicator objects which are then run to perform the communications. 
Fields can therefore be grouped into several communicator objects, performing communication at different times.

Domain decomposition is in two dimensions ($x$ and $y$), and is currently done as a regular grid: the number
of points in each dimension is the same on each processor. Decomposition in $z$ is a possible future extension, but is complicated
because FFTs are used in this dimension for Laplacian inversion (section~\ref{sec:operators}) and 
the shifting needed in tokamak field-aligned coordinates (section~\ref{sec:tok_coord}).

Topology is handled internally in the communication object, using branch-cuts specified in the grid input file.
This is important in, for example, simulations of tokamaks in x-point geometry
where the domain is not topologically rectangular. Within each processor's 
domain the grid is topologically rectangular, simplifying differencing methods, so branch-cuts must coincide
with processor boundaries.

\subsection{Coordinate system}
\label{sec:coords}

Coordinate systems are implemented by using global scalar field objects
for each component of the metric tensors $g^{ij}$ and $g_{ij}$, and Christoffel
symbol $\Gamma^i_{jk}$ components calculated from these. All differential operators
(section~\ref{sec:operators}) then use these quantities.

When a grid file is loaded, these quantities are read if they are present,
otherwise they can be set in the physics module. Since metric tensor quantities are not
fixed, this could in principle be used to implement moving meshes by evolving the metric tensor, 
although this has not yet been attempted.

Currently the coordinate system is restricted to having one symmetry direction ($z$), so that the metric
tensor components are 2D fields. Changing this to allow fully 3D metric tensors would be straightforward
(and is planned as a future option), but is not currently necessary for tokamak simulations.

\subsubsection{Tokamak coordinate systems}
\label{sec:tok_coord}

An important class of instabilities in tokamaks produces structures which are highly elongated along 
magnetic field-lines ($k_{||} << k_\perp$, where $k_{||}$ and $k_\perp$ are wavenumbers parallel and
perpendicular to the magnetic field respectively). Aligning the computational mesh along the magnetic field
therefore allows far fewer grid-points to be used in this direction, with a corresponding reduction
in the computational cost of a simulation. Due to the periodicity and complications introduced by
magnetic shear, some special features have been implemented to handle these coordinate systems
which are discussed here.

\begin{figure}[htb!]
\centering
\includegraphics[scale=0.4]{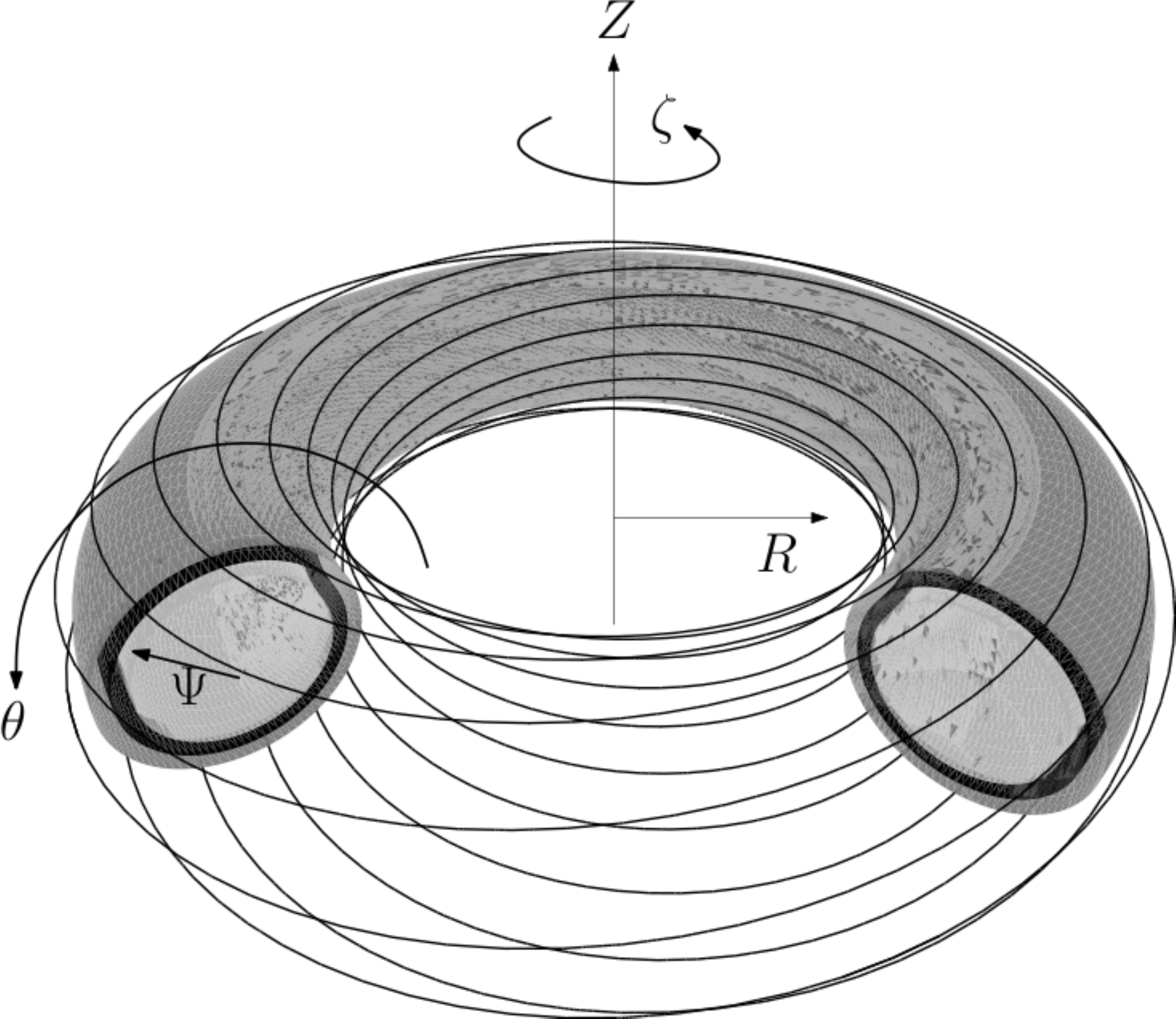}
\caption{Diagram of field-lines and flux-surfaces in a tokamak. Major radius $R$, height $Z$, poloidal flux $\Psi$, 
poloidal angle $\theta$, and toroidal angle $\zeta$.}
\label{fig:tokamak}
\end{figure}
The derivation of the field-aligned coordinate system starts with an orthogonal toroidal
coordinate system $\left(\psi, \theta, \zeta\right)$, illustrated in figure~\ref{fig:tokamak}.
$\psi$ is the poloidal flux, $\theta$ the poloidal angle (from $0$ to $2\pi$), and $\zeta$ the
toroidal angle (also $0$ to $2\pi$).

Aligning the mesh with the equilibrium magnetic field, grid-points are placed in a coordinate system
defined by \cite{bout_manual}:
\begin{equation}
x = \psi - \psi_0 \qquad y = \theta \qquad z = \zeta - \int_{\theta_0}^{\theta}\nu\left(\psi, \theta\right)d\theta
\label{eq:coordtransform}
\end{equation}
Where $\nu$ is the local field-line pitch given by
\begin{equation}
\nu\left(\psi, \theta\right) = \frac{\mathbf{B}\cdot\nabla\zeta}{\mathbf{B}\cdot\nabla\theta} = \frac{\Bt\hthe}{\Bp R}
\end{equation}
where $h_\theta = 1 / \left|\nabla\theta\right|$ is the $\theta$ scale factor. 
The contravariant basis vectors are therefore
\begin{eqnarray*}
\nabla x &=& \nabla \psi \qquad \nabla y = \nabla \theta \\
\nabla z &=& \nabla\zeta - I \nabla\psi - \nu\left(\psi, \theta\right)\nabla\theta
\end{eqnarray*}
where $I = \int_{\theta_0}^\theta\partial\nu\left(\psi, \theta\right) / \partial\psi d\theta$ is the integrated local shear.
Physically, different flux surfaces are labelled by $x$, while different field lines on a flux surface are labelled
by $z$ (i.e. $\mathbf{B}\cdot\nabla z = 0$).
The covariant basis vector (the vector between grid-points) is therefore:
\begin{eqnarray}
\mathbf{e}_x &=& \frac{1}{RB_\theta}\mathbf{\hat{e}}_\psi + IR\mathbf{\hat{e}}_\zeta \label{eq:cov_basis} \\
\mathbf{e}_y &=& \frac{h_\theta}{B_\theta}\mathbf{B} = h_\theta\mathbf{\hat{e}}_\theta + \nu R\mathbf{\hat{e}}_\zeta \nonumber \\
\mathbf{e}_z &=& R\mathbf{\hat{e}}_\zeta \nonumber
\end{eqnarray}
where $\mathbf{\hat{e}}$ are the unit vectors in the original orthogonal toroidal $\left(\psi, \theta, \zeta\right)$
coordinate system.

The $\theta$ periodicity now becomes a boundary condition on $y$: after a full poloidal circuit
the mesh has now been shifted toroidally by $2\pi q$ radians where $q\left(\psi\right)=\left(1/2\pi\right)\oint\nu d\theta$
is the standard ``safety factor'' \cite{wesson-1997}. This
shifted mesh must then be mapped onto the original mesh using toroidal periodicity at a fixed
poloidal location (called the twist-shift condition \cite{bout_manual}). 

The twist-shift condition guarantees $\theta$ (poloidal) periodicity of field values on the grid, but not
of radial derivatives. This is due to the magnetic shear, the variation in safety factor with
flux-surface: following a bundle of field-lines around the torus, it becomes sheared as the field-lines
on one surface have a different pitch to those on another. When this field-line bundle completes one
poloidal circuit
of the torus and is connected back onto itself (periodicity constraint), shear produces a discontinuity
in the radial derivative. 
\begin{figure}[htb!]
\centering
\includegraphics[scale=0.5]{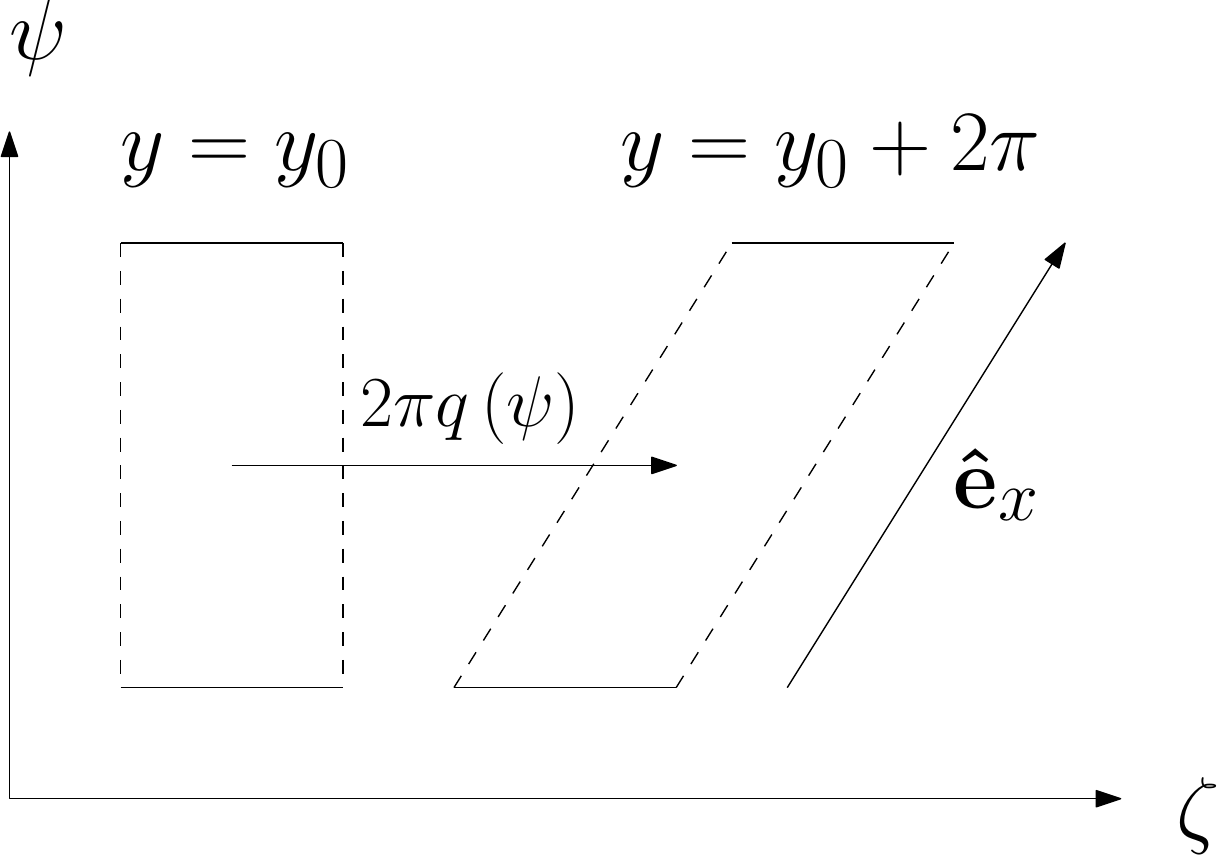}
\caption{Radial derivatives in a field-aligned coordinate system, showing how 4 mesh points
map as one travels once around the $\theta$ (poloidal) direction}
\label{fig:shear}
\end{figure}
This can be seen in the radial covariant basis vector
$\mathbf{e}_x$ (equation~\ref{eq:cov_basis}) and illustrated in figure~\ref{fig:shear}: at $y=y_0$, $I = 0$
and $\mathbf{e}_x$ is
in the $\nabla\psi$ direction, whereas at $y = y_0 + 2\pi$, $I\neq 0$ and so the coordinate 
system has a discontinuity. For differencing methods this corresponds to using a stencil
which is discontinuous in space across this matching location $y_0$.  This ``special'' poloidal
location $y_0$ is unphysical, and is often the source of numerical instability.

The solution to this problem which is used in BOUT++ for tokamak simulations (section~\ref{sec:elmsim}),
is to use ``quasi-ballooning'' coordinates given in \cite{dimits-1993}, similar to the
scheme used by the GEM codes \cite{scott-2005}. 
Calculation of differentials are performed in a patchwork of local coordinate systems, in which the $\mathbf{e}_x$ 
basis vector is in the $\nabla\psi$ direction i.e. the $I$ term in equation~\ref{eq:cov_basis} is dropped. This
coordinate system no longer contains a ``special'' poloidal location, but instead flux-surfaces
slide past each other. In general, grid-points will not be aligned in the $\nabla\psi$ direction,
and so interpolation in the toroidal ($\zeta$) direction is required to perform 
$x$ derivatives. Since the domain is periodic in this $z$, FFTs are used to perform this interpolation.

Existing codes employing this type of shifted coordinate system solve scalar equations such
as reduced MHD or gyro-fluid (in the case of GEM) equations. A complication arises however 
for vector equations because the local coordinate system is non-commutative, and so cannot be fully
described by a metric tensor: moving in the $\mathbf{e}_\psi$ direction and then
along the $\mathbf{e}_y$ ($\mathbf{B}$ field) direction is different to moving in $\mathbf{e}_y$
then $\mathbf{e}_\psi$ directions, due to magnetic shear.
In differential geometry this is called a coordinate system torsion \cite{lovelock-1989}. In this case there is a change
in the curl operator, which must now include a term due to this shifting between coordinate systems, 
proportional to the magnetic shear:
\begin{equation}
\Curl{\mathbf{A}} \rightarrow \Curl{\mathbf{A}} + \frac{1}{\sqrt{g}}\deriv{\nu}{\psi} A_z \mathbf{e}_z
\end{equation}
i.e. 
\[
\left(\Curl{A}\right)^z = \frac{1}{\sqrt{g}}\left(\deriv{A_y}{\psi} - \deriv{A_\psi}{y}\right) + \frac{1}{\sqrt{g}} \deriv{\nu}{\psi} A_z
\]
Note that this term is not a physical effect, but is an artifact of using a non-commutative set of coordinates
for differencing. In most tokamak simulations this term is expected to be small, but should be considered. 

Both the ballooning $\left(x,y,z\right)$ coordinates, and shifted quasi-ballooning $\left(\psi, y, z\right)$ 
coordinates have problems in handling magnetic shear. Quasi-ballooning coordinates are used in BOUT++ 
simulations (though both systems can be employed) because non-commutativity is preferable to
the introduction of a special poloidal location and numerical instability.

\subsection{Differencing methods}
\label{sec:difops}

BOUT++ is a finite-differencing code using Method of Lines (MOL), so that differentials
are calculated in each dimension separately. Because of this, differential operators can be
split into two components: the calculation of partial derivatives
(e.g. $\partial/\partial x$) on the grid, and the use of these functions to implement
differential operators using a specified metric tensor.

The choice of differencing methods to be used is quite problem-specific, and so can be changed
at run-time in the input file. Methods can be specified for central derivatives
(first and second derivatives), and upwinding in each dimension separately.
Currently the methods implemented for central derivatives are
$2^{nd}$ order, $4^{th}$ order, Central Weighted Essentially Non Oscillatory (CWENO) \cite{levy-1999,bryson-2003}
and derivatives using the Fast Fourier Transform (FFT) in the $z$ (axisymmetric) dimension.

Advection terms require special treatment and so the following schemes are currently
implemented: first order upwinding, and 3rd-order 
Weighted Essentially Non Oscillatory (WENO) \cite{jiang-1996,jiang-1997}.
WENO methods provide high accuracy, whilst remaining well-behaved at steep
gradients such as shocks, and this scheme has been used in all the tests presented
in section~\ref{sec:tests} and ELM simulations in section~\ref{sec:elmsim}.

\subsubsection{Operators}
\label{sec:operators}

Differential operators use the differencing methods specified in the input option file, 
and metric tensor components from the grid input file. Operators
are divided into two classes: those which are independent of any coordinate system, and those which are
intended for use in a Clebsch coordinate system where $\mathbf{B} = \nabla x \times \nabla z$, i.e with 
$\mathbf{B}$ aligned with the $y$ coordinate. Because different numerical methods are needed for upwinding
terms, the operation $\mathbf{v}\cdot\nabla f$ has a special function \code{V\_dot\_Grad(v, f)}. 

\[
\begin{array}{rclrcl}
\mathbf{v} &=& \nabla f &\qquad \code{Vector} &=& \code{Grad(Scalar)} \\
f &=& \nabla\cdot\mathbf{a} &\qquad \code{Scalar} &=& \code{Div(Vector)} \\
\mathbf{v} &=& \nabla\times\mathbf{a} &\qquad \code{Vector} &=& \code{Curl(Vector)} \\
f &=& \mathbf{v}\cdot\nabla g &\qquad \code{Scalar} &=& \code{V\_dot\_Grad(Vector, Scalar)} \\
\mathbf{v} &=& \mathbf{a}\cdot\nabla\mathbf{b} &\qquad \code{Vector} &=& \code{V\_dot\_Grad(Vector, Vector)} \\
f &=& \nabla^2 f &\qquad \code{Scalar} &=& \code{Laplacian(Scalar)}
\end{array}
\]

These are operators which assume that the equilibrium magnetic field is written
in Clebsch form as
\[
\mathbf{B}_0 = \nabla z\times\nabla x \qquad \left|B_0\right| = \frac{\sqrt{g_{yy}}}{J}
\]
These operators include:
\[
\begin{array}{rclrcl}
\partial^0_{||} &=& \mathbf{b}_0\cdot\nabla &\qquad \code{Scalar} &=& \code{Grad\_par(Scalar)} \\
\nabla^0_{||}F &=& B_0\partial^0_{||}\left(\frac{F}{B_0}\right) &\qquad \code{Scalar} &=& \code{Div\_par(Scalar)} \\
f &=& \mathbf{b}_0\cdot\nabla\phi\times\nabla A &\qquad \code{Scalar} &=& \\
  & & \multicolumn{4}{c}{\code{b0xGrad\_dot\_Grad(Scalar, Scalar)}}
\end{array}
\]

A common problem encountered in plasma fluid simulations is to invert an equation of the form:
\[
\nabla_\perp^2 x + a x = b
\]
With the operator $\nabla_\perp = \nabla - \mathbf{b}\left(\mathbf{b}\cdot\nabla\right) = -\mathbf{b\times} \left(\mathbf{b\times}\nabla\right)$.
This operator appears in reduced MHD for the vorticity inversion, and is used in many
of the tests described in section~\ref{sec:tests}, and ELM simulations in
section~\ref{sec:elmsim}. 
Efficiently inverting this operator is done in the same way as in the BOUT code:
FFTs are used in the $z$ direction to transform this problem into a set of 1D 
inversion problems (in $x$) for each Fourier mode.
These inversion problems are band-diagonal (tri-diagonal in the case of 2nd-order differencing) and so
inversions are very efficient: $O\left(n_z \log n_z\right)$ for the FFTs, $O\left(n_x\right)$ for tridiagonal inversion
using the Thomas algorithm \cite{press-1999}, where $n_x$ and $n_z$ are the number of grid-points
in the $x$ and $z$ directions respectively.

\subsection{Physics module}
\label{sec:physics_model}

This module determines the actual equations solved by BOUT++, and is the only
part of BOUT++ which `knows' what the variables physically mean.
Physics modules have to implement two functions: \code{physics\_init},
which is called once at the start of the run and initialises variables,
and \code{physics\_run} which is called every time-step, and
calculates time-derivatives for a given state (see figure~\ref{fig:pvode},
section~\ref{sec:time_int}). To illustrate the clarity possible with BOUT++,
the equations of ideal MHD and the corresponding lines
of code are shown in table~\ref{tab:mhdcode}.

\begin{table}[htbp!]
\caption{Comparison of analytic Ideal MHD expressions, and the corresponding BOUT++ code}
\label{tab:mhdcode}
\centering
\begin{tabular}[c]{ r  l  r l }
  \hline
 \multicolumn{2}{c}{Analytic} & \multicolumn{2}{c}{BOUT++} \\
\hline
$\partial n / \partial t = $ & & \texttt{Scalar3D } & \texttt{dndt =} \\
& $-\mathbf{v}\cdot\nabla n$ & & \texttt{- V\_dot\_Grad(v, n)} \\
& $- n\nabla\cdot\mathbf{v}$ & & \texttt{- n*Div(v)} \\
$\partial p / \partial t =$ & & \texttt{Scalar3D } & \texttt{dpdt =} \\
& $-\mathbf{v}\cdot\nabla p$ & & \texttt{- V\_dot\_Grad(v, p)} \\
& $- \gamma p\nabla\cdot\mathbf{v}$ & & \texttt{- gamma*p*Div(v)} \\
$\partial \mathbf{v} / \partial t =$ & & \texttt{Vector3D } & \texttt{dvdt =} \\
& $ -\mathbf{v}\cdot\nabla\mathbf{v}$ & & \texttt{- V\_dot\_Grad(v, v)} \\
& $-\nabla p / n$ & & \texttt{- Grad(p)/n} \\
& $+\frac{1}{n}\left(\nabla\times\mathbf{B}\right)\times\mathbf{B}$ & & \texttt{+ (1/n)*(Curl(B)\pow B)} \\
$\partial \mathbf{B} / \partial t =$ & & \texttt{Vector3D } & \texttt{dBdt =} \\
& $\nabla\times\left(\mathbf{v}\times\mathbf{B}\right)$ & & \texttt{Curl(v\pow B)} \\
\hline
\end{tabular}
\end{table}

In addition to the evolution equations, some initialisation code is needed to set up the simulation problem. 
This initialisation function \code{physics\_init} consists of
\begin{itemize}
\item Definition of fields to store state and time-derivatives (declared as global variables)
\item Loading initial profiles
\item Calls to specify which fields to use for state and time-derivatives. 
\item Creation of a communications object, and specification of the fields to communicate (optional)
\item Addition of extra variables to the output files (optional)
\end{itemize}

An important component of the problem specification is the boundary conditions. In BOUT++, the
boundary conditions for each evolving variable can be set in the input settings file (section~\ref{sec:io}),
allowing the effect of changing boundary conditions to be quickly assessed. Currently, possible
boundary conditions on a scalar field $f$ include zero-value, zero-gradient, $\delp f = 0$, (anti-) symmetric.
Generalised implementation of more complicated coupled boundary conditions is a possible future extension.

Putting all the problem-specific code in one place allows a user to quickly verify the equations
being solved, and to quickly implement new physical models. In the next section
test problems using a variety of physics modules are presented.

\section{Test problems}
\label{sec:tests}

Three test problems are presented here, the first two of which 
were published in \cite{umansky-2008-tests}, and have also been used
to benchmark the BOUT code \cite{umansky-2006-bout}: the resistive drift
instability (section~\ref{sec:drift})
tests the fidelity with which the code simulates wave propagation and in particular wave phase shifts.
An interchange mode in a curved slab (section~\ref{sec:interchange}) is a simplified form of the
ballooning mode, and so recovering the growth-rate of this mode is important for the later ELM simulations.
Finally, the Orszag-Tang vortex problem in ideal MHD is simulated in section~\ref{sec:otv}. This tests the
numerical stability and accuracy of BOUT++ in simulating shocks, which is potentially important for non-linear ELM
simulations.

\subsection{Resistive drift-wave instability}
\label{sec:drift}

A drift-wave is a wave which exists in a plasma wherever there is a pressure
gradient \cite{hazeltine-2003}.
Without dissipation, fluctuations in density $n$ and electrostatic potential $\phi$ are in phase
so there is no transport of plasma and the wave amplitude does not grow.
Dissipation, in this case resistivity, introduces a phase-shift
between $n$ and $\phi$ and hence transport of plasma and growth of the mode.
Since all that is required for radial transport is a pressure gradient
and some form of dissipation (in the absence of magnetic shear), this is often called the ``universal''
instability. 
Because the growth of the resistive drift-wave instability is sensitive to
phase shifts, this test checks how accurately this phase is simulated.

The equations solved are for the density $n$, and vorticity
$\omega = n_0\mathbf{b}\cdot\Curl{\mathbf{v}}$.
The simulation is electrostatic, and the zero electron mass approximation
is used to obtain the parallel current $j_{||}$. All quantities with a '$0$'
subscript are equilibrium and not evolved.
\begin{eqnarray*}
\deriv{n}{t} &=& -\mathbf{V}_E\cdot\nabla n_0 \\
\deriv{\omega}{t} &=& \frac{B_0^2}{m_i}\nabla_{||}j_{||} \\
\mathbf{V}_E &=& \frac{1}{B_0}\mathbf{b}_0\times\nabla_\perp\phi \\
\nabla_\perp^2\phi &=& \omega/n_0 \\
j_{||} &=& \sigma_{||}\left(T_0\partial_{||}n - n_0\partial_{||}\phi\right)
\end{eqnarray*}
The simulation domain is a cylindrical annulus with radius $R = 5.4$~m, radial width $6$~cm and constant
density scale-length $L_N = 4.5$~m. This is a 2D periodic simulation domain, but since perpendicular
wavenumber is fixed in a given simulation, the simulation is effectively 1D. Radial boundary conditions are
zero-gradient vorticity and density, and $\phi = 0$.

The analytic dispersion relation is $\left(\omega - \omega_*\right)i\sigma_{||} + \omega^2 = 0$, with
diamagnetic frequency $\omega_* = k T_{e0} / L_N$ \cite{umansky-2008-tests}.
\begin{figure}[htbp!]
\centering
\subfigure[Growth rate]{
  \label{fig:drift_imag}
  \includegraphics[scale=0.35]{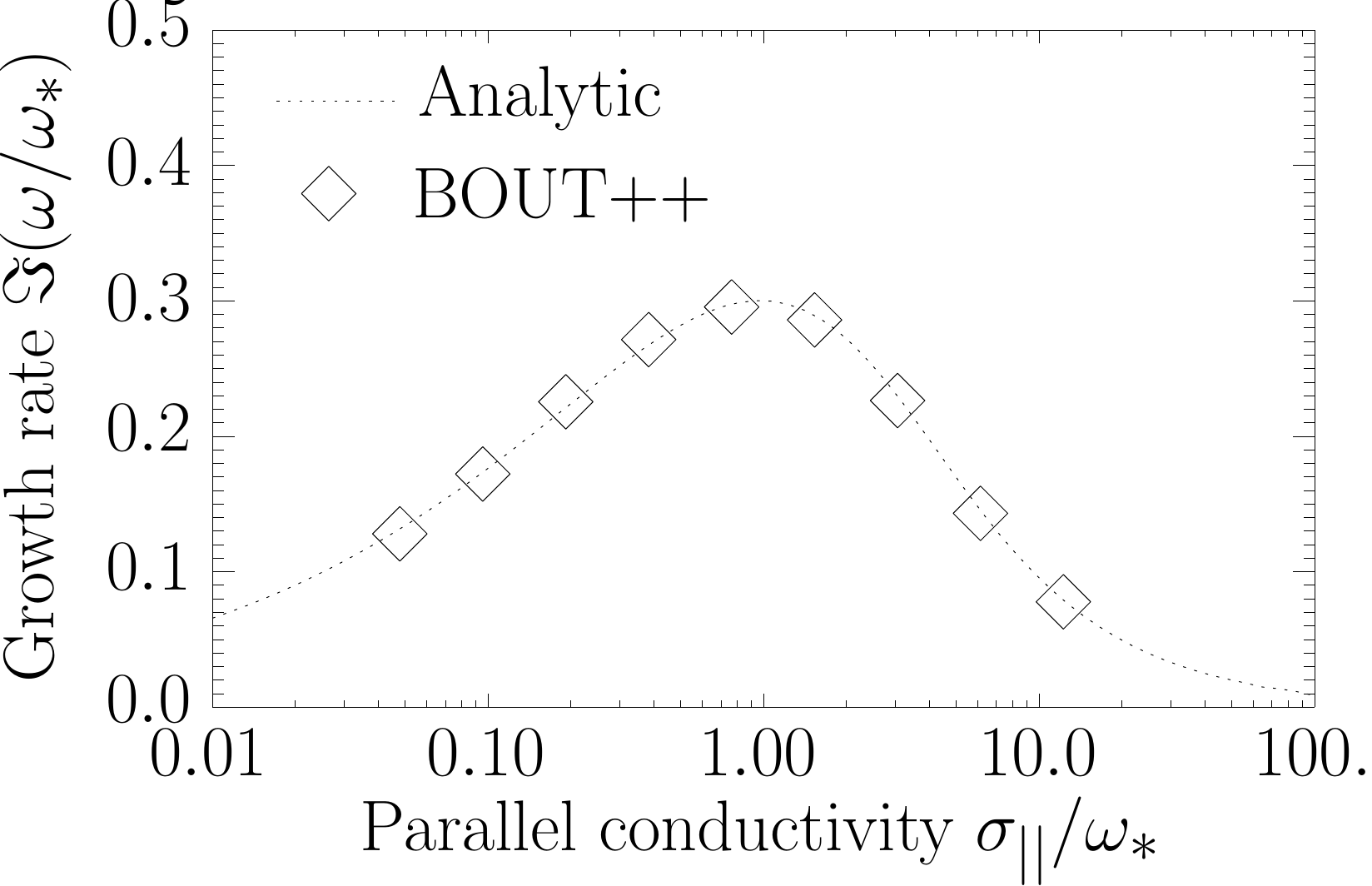}
}
\subfigure[Real Frequency]{
  \label{fig:drift_real}
  \includegraphics[scale=0.35]{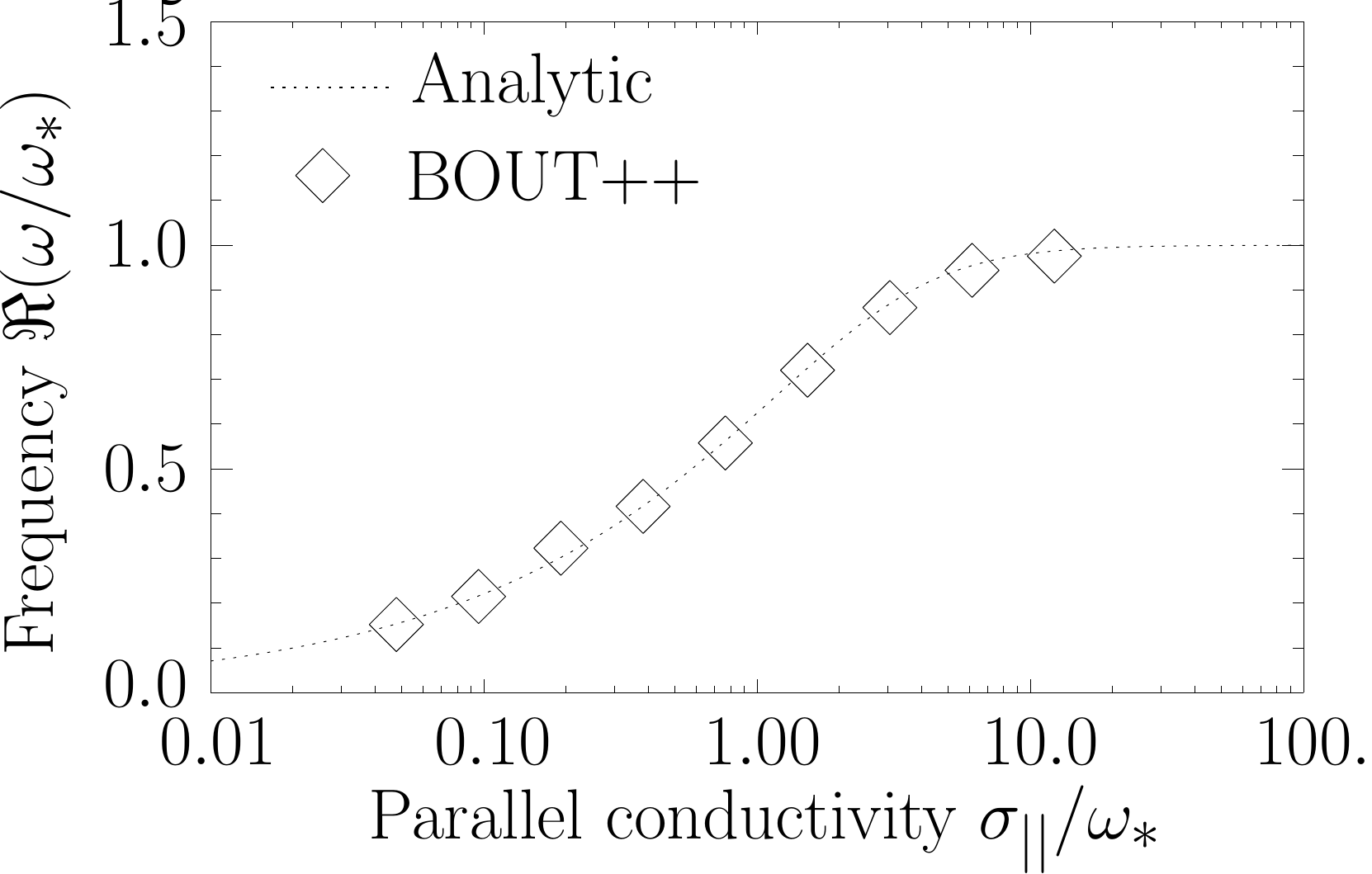}
}
\caption{Resistive Drift wave instability test. Dashed lines are analytic results, diamonds from BOUT++ simulations}
\label{fig:drift_test}
\end{figure}
Figure~\ref{fig:drift_test} shows the analytic growth rate and real frequency for 
this mode (dashed line), and the BOUT++ results (diamonds). As the parallel conductivity
$\sigma_{||}$ is varied there is a peak in the growth-rate, the location of
which is recovered well by BOUT++, indicating
that wave phases are accurately simulated.

\subsection{Interchange mode}
\label{sec:interchange}

The interchange mode is an instability driven by pressure gradients and curvature, and has some
features in common with a ballooning mode. This test is therefore a highly simplified version
of the ELM problem simulated in section~\ref{sec:elmsim}.

The simulation domain is a curved slab with radius of curvature $R$, periodic in $z$ and with zero-gradient
boundary conditions in $x$ and $y$ \cite{umansky-2008-tests}. As with the drift instability test, 
this domain is 2D, but the wave-number is fixed in one of these dimensions. The equations
solved are for density $n$ and vorticity $\omega$: 
\begin{eqnarray*}
\deriv{n}{t} &=& -\mathbf{V}_E\cdot\nabla n_0 \\
\deriv{\omega}{t} &=& 2\omega_{ci}\left(\mathbf{b}_0\times\kappa_0\right)\cdot\nabla p \\
\mathbf{V}_E &=& \mathbf{b}_0\times \nabla_\perp \phi/B \\
q\nabla_\perp^2\phi &=& \omega/n_0 \\
p &=& 2T_0n
\end{eqnarray*}
with the magnetic field curvature vector
$\mathbf{\kappa_0} = \mathbf{b}_0\cdot\nabla\mathbf{b}_0 \sim 1 / R$. The density gradient is in the $x$ direction
with a length-scale of $2$~cm, and the temperature $T_0$ is a constant.

\begin{figure}[htb!]
\centering
\includegraphics[scale=0.4]{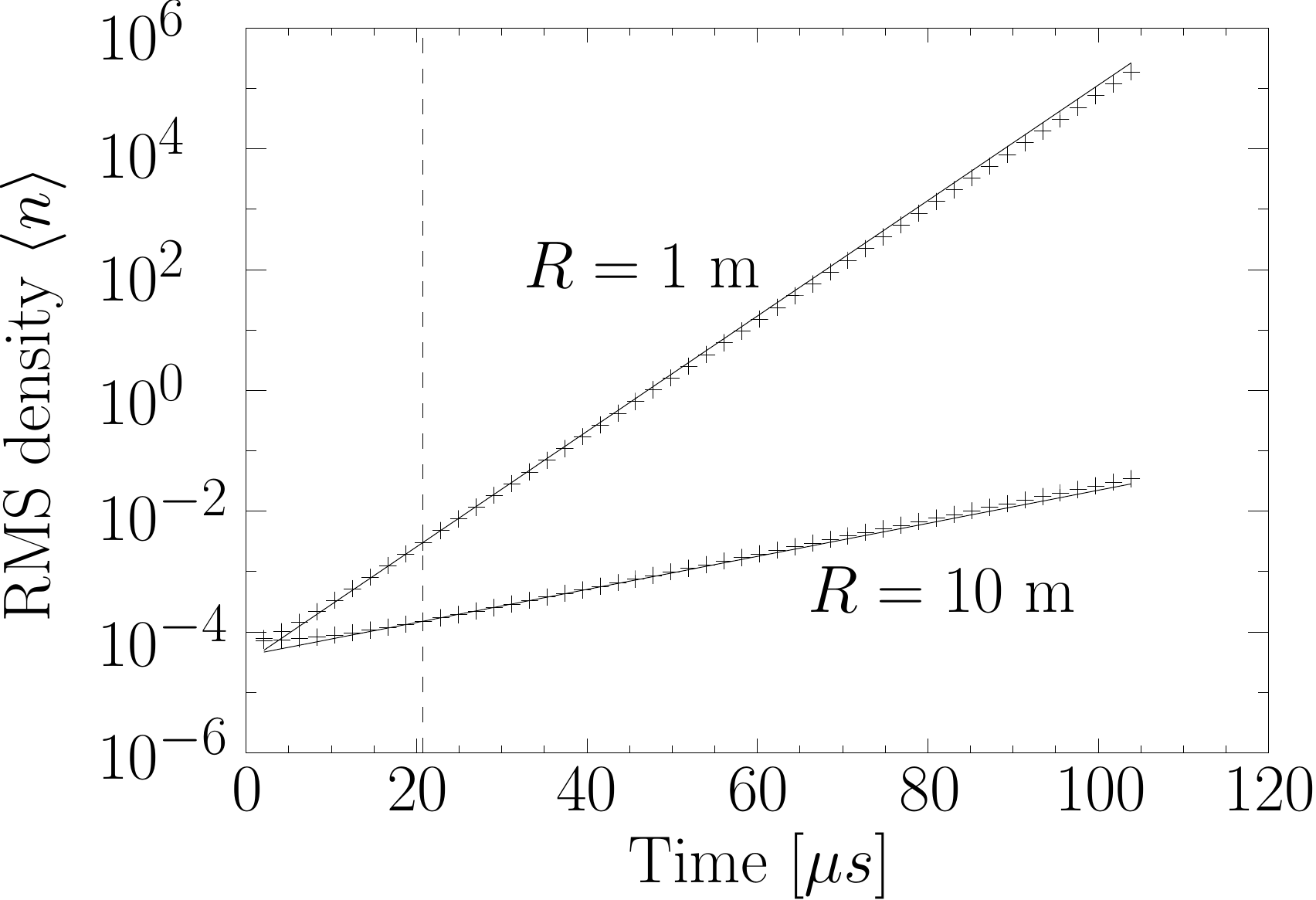}
\caption{Interchange instability test. Solid lines are from analytic theory, symbols from BOUT++ simulations, and the RMS
density is averaged over $z$. Vertical dashed line marks the reference point, where analytic and simulation results are set equal}
\label{fig:profiles}
\end{figure}
Figure~\ref{fig:profiles} shows the time-dependence of density for two cases with $R=1$ and $10$ metres. This
shows that the growth-rate (slope of each line) is well reproduced in both cases, giving some confidence in the
simulation of ELMs to be discussed in section~\ref{sec:elmsim}. In addition, this growth-rate is maintained over a long period (mode amplitude increases by
8 orders of magnitude in the case of $R=1$~m) without noise or other numerical problems significantly affecting the result.

\subsection{Orszag-Tang vortex}
\label{sec:otv}

This is a standard test problem for multi-dimensional MHD codes which tests how robust a
scheme is to the formation of MHD shocks, and the accuracy with which the $\nabla\cdot\mathbf{B}=0$
condition is preserved \cite{orszag-1979,dai-1998}. The equations solved are ideal MHD:
\begin{eqnarray*}
\deriv{n}{t} &=& -\mathbf{v}\cdot\nabla n - n\nabla\cdot\mathbf{v} \\
\deriv{p}{t} &=&  -\mathbf{v}\cdot\nabla p - \gamma p\nabla\cdot\mathbf{v} \\
\deriv{\mathbf{v}}{t} &=& -\mathbf{v}\cdot\nabla\mathbf{v} + \frac{1}{n}\left[\left(\nabla\times\mathbf{B}\right)\times\mathbf{B} - \nabla p\right] \\
\deriv{\mathbf{B}}{t} &=& \nabla\times\left(\mathbf{v}\times\mathbf{B}\right) \\
\end{eqnarray*}
in a periodic 2D domain with sides of length 1. Mass density $\rho = 25/(36\pi)$
and pressure $p = 5/(12\pi)$ are uniform (with sound speed $C_S = 1$), and the initial conditions for velocity and magnetic
field are:
\begin{eqnarray}
\mathbf{v}_0\left(x, y\right) = \left[-sin\left(2\pi y\right), sin\left(2\pi x\right)\right] \nonumber \\
\mathbf{B}_0\left(x, y\right) = \frac{1}{\sqrt{4\pi}}\left[-sin\left(2\pi y\right), sin\left(4\pi x\right)\right] \nonumber
\end{eqnarray}

\begin{figure}[htbp!]
\centering
\subfigure[Pressure at $t = 0.5$]{
  \label{fig:ot_p}
  \includegraphics[scale=0.50]{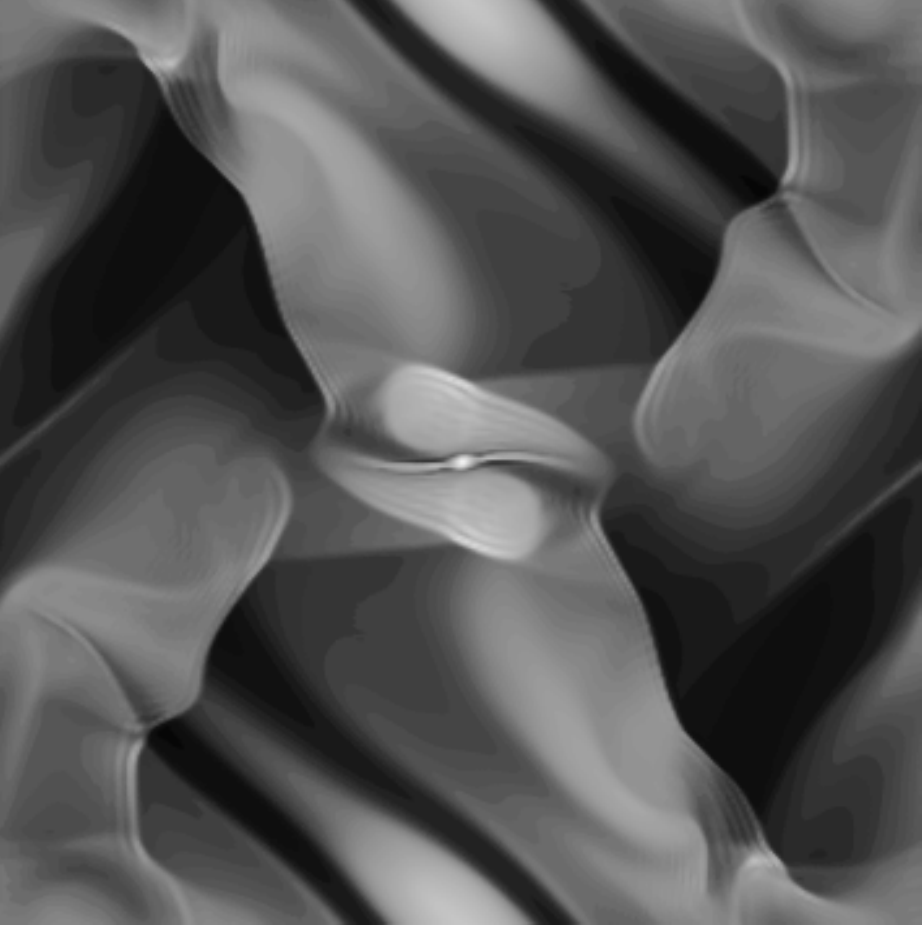}
}
\subfigure[Divergence of B: maximum (solid) and RMS (dashed)]{
  \label{fig:ot_divb}
  \includegraphics[scale=0.35]{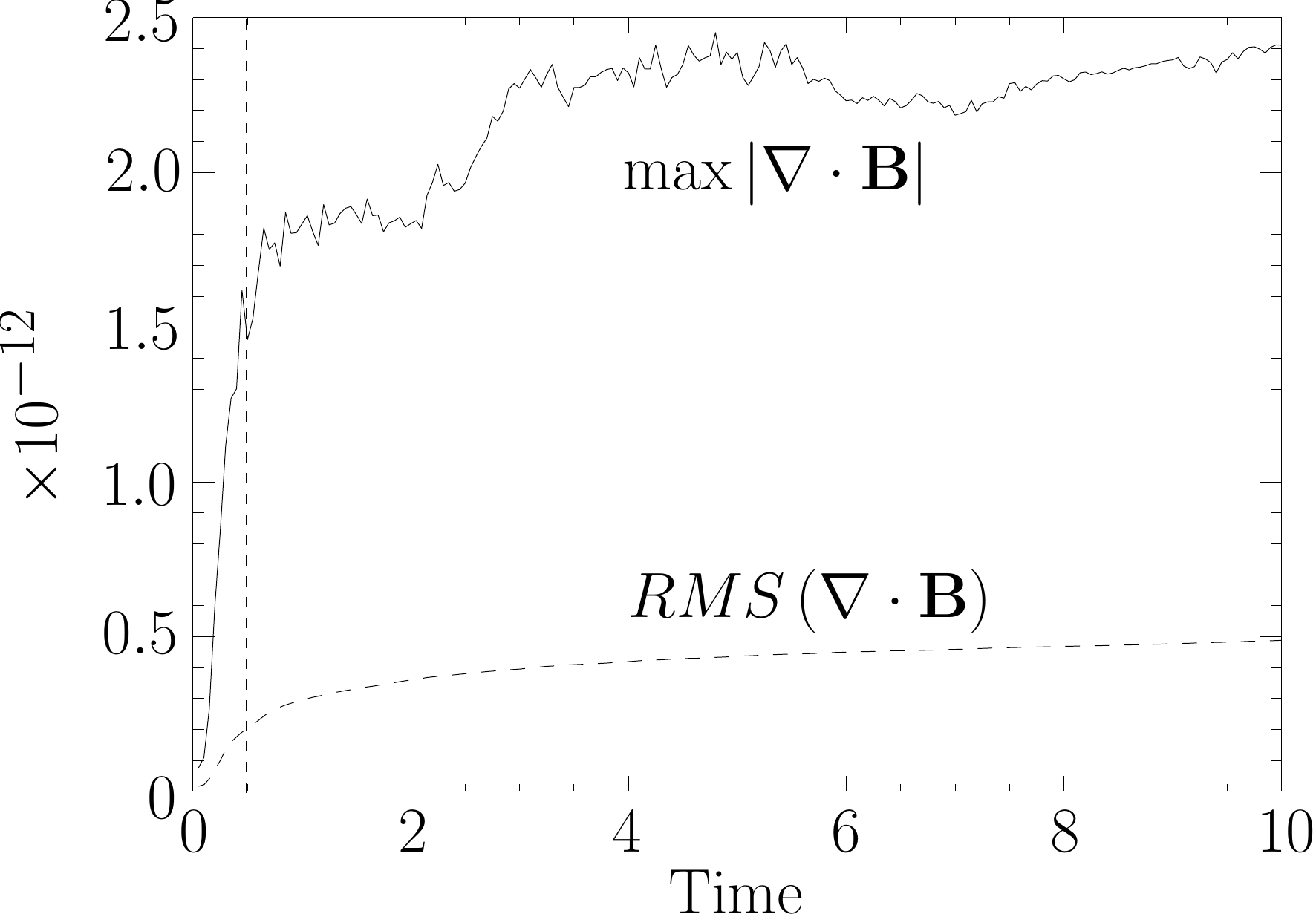}
}
\caption{Orszag-Tang vortex test on a $256 \times 256$ grid}
\label{fig:ot_test}
\end{figure}

The simulation results at $t=0.5$ shown in figure~\ref{fig:ot_p} agrees
qualitatively with those from ideal MHD codes such as ATHENA \cite{stone-2008}.
The divergence of $\mathbf{B}$ is shown in figure~\ref{fig:ot_divb}, indicating that
the formation of shocks leads to an increase in $\left|\nabla\cdot\mathbf{B}\right|$. 
At the time shown in figure~\ref{fig:ot_p}
$t=0.5$ (vertical dashed line in figure~\ref{fig:ot_divb}),
$\left|\nabla\cdot\mathbf{B}\right| = 1.5\times 10^{-12}$ for a $128\times 128$ mesh,
$4.6\times 10^{-12}$ on a $256 \times 256$ grid and $1.8\times 10^{-11}$ on a $512 \times 512$ grid.
The large-scale $B / L$ values are $\sim 10^{-1}$, demonstrating that the numerical
methods used can maintain the $\nabla\cdot\mathbf{B} = 0$ condition to high accuracy. 

Note that the simulation progresses well beyond the point shown in figure~\ref{fig:ot_p}
as can be seen in figure~\ref{fig:ot_divb} where the simulation runs to $t=10$, with only very slow increase
in $\nabla\cdot\mathbf{B}$.

\begin{figure}[htb!]
\centering
\includegraphics[scale=0.4]{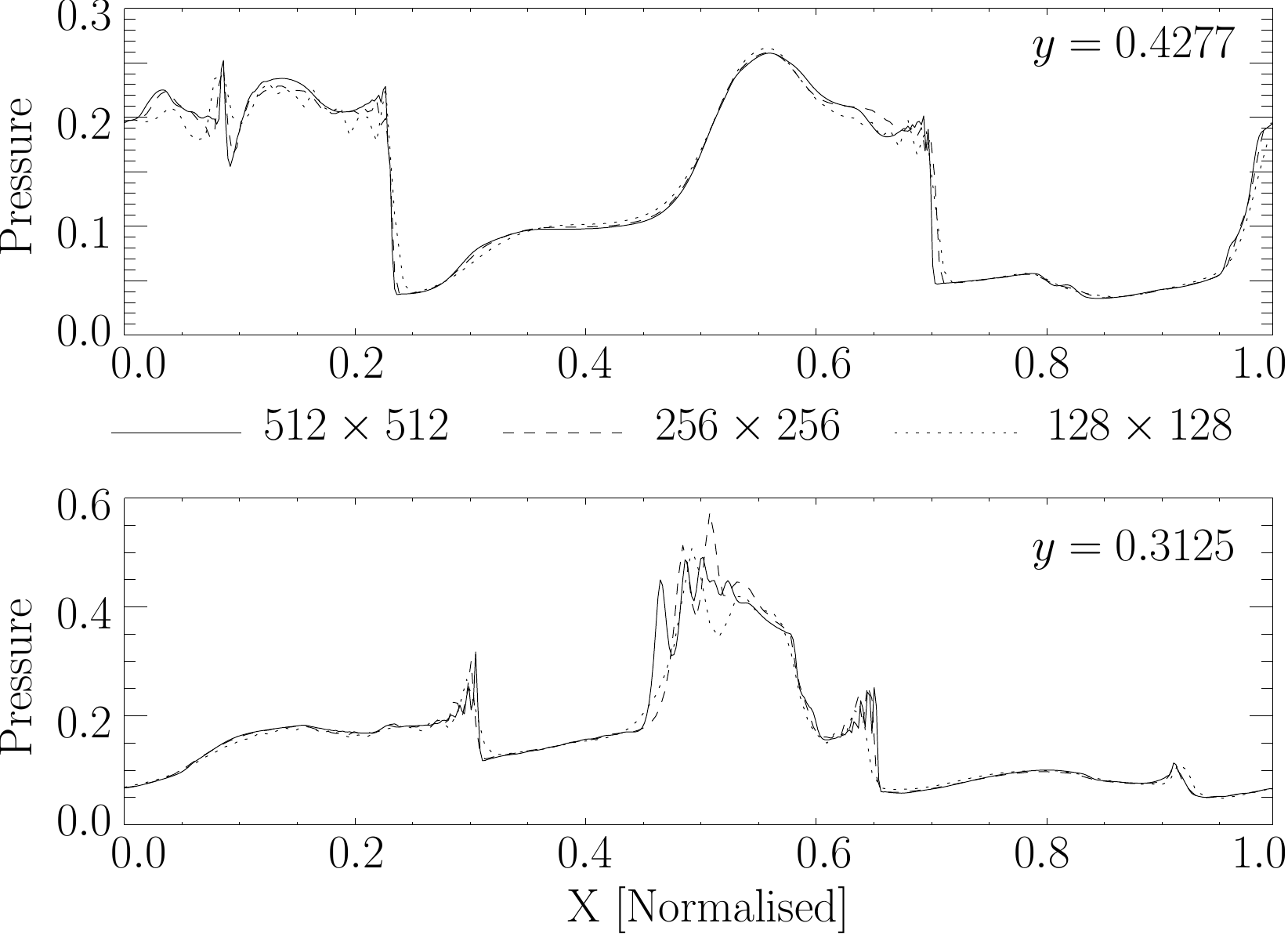}
\caption{Slices through the Orszag-Tang vortex solution for three different grid resolutions}
\label{fig:otv_yslices}
\end{figure}

A more quantitative comparison with other codes is to take slices through this solution at
$y=0.3125$ and $y=0.4277$, shown in figure~\ref{fig:otv_yslices}.
These can be compared with figure~11 in \cite{londrillo-2000}, figure~3 in \cite{ryu-1998},
and figure~9 in \cite{zhang-2006}. 
Figure~\ref{fig:otv_yslices} shows that whilst the BOUT++ solutions are
very close to those from specialised MHD codes, it is susceptible to
oscillations at the top of shocks.
These oscillations do not grow as the
simulation progresses, and improve as grid resolution is increased.
They are present because many terms in the equations are
solved using central methods ($4^{th}$-order central differencing) - only the upwinding terms
use WENO type methods for differencing, and currently no flux-splitting is employed. 

Handling of shocks and sharp gradients in BOUT++ is currently acceptable, in that
they do not produce numerical instabilities or unphysical values: the WENO advection
maintains positive-definite density and pressure, and the solution in smooth regions is accurate.
Future work includes improving the handling of shocks to remove the small unphysical
oscillations observed in figure~\ref{fig:otv_yslices}.

\section{ELM simulations}
\label{sec:elmsim}

The primary motivation for developing this code is the simulation of Edge Localised Modes (ELMs)
in tokamaks (see section~\ref{sec:phys_overview}). 
In this section we present linear benchmarking of BOUT++, 
comparing the results with the ELITE linear MHD eigenvalue code \cite{wilson-2002,snyder-2002}.
The equations solved using BOUT++ are high-$\beta$ reduced ideal MHD \cite{hazeltine-2003}, evolving
vorticity $\omega=\mathbf{b}\cdot\Curl{\mathbf{v}}$, pressure $p$, and parallel component of the vector potential 
$\apar = \bvec\cdot\mathbf{A}$ :
\begin{eqnarray*}
\rho_0 \frac{d\omega}{dt} &=& B_0^2\bvec\cdot\nabla\left(\frac{\Jpar}{B_0}\right) + 2\bxk p \\
\deriv{\apar}{t} &=& -\nabla_{||}\phi \\
\frac{dp}{dt} &=& -\frac{1}{B_0}\bvec_0\times\nabla\phi\cdot\nabla p_0 \\
\omega &=& \frac{1}{B_0}\delp\phi \\
\Jpar &=& J_{||0} - \frac{1}{\mu_0}\delp\apar \\
\frac{d}{dt} &=& \deriv{}{t} + \frac{1}{B_0}\bvec_0\times\nabla\phi\cdot\nabla \\
\end{eqnarray*}
where $\rho_0$ is the mass density (which is a constant); $J_{||} = \mathbf{b\cdot J}$ the parallel current; $\phi$ the electrostatic
potential; $\kappa_0 = \mathbf{b}_0\cdot\nabla\mathbf{b}_0$ is the field curvature
(as for the interchange test, section~\ref{sec:interchange}), and everything is in SI units. 
The perturbed magnetic field is given in terms of the parallel vector potential by
$\mathbf{B}_1 =  \nabla\apar\times\mathbf{b}_0$. The vorticity equation includes
the kink/peeling term through the perturbed magnetic field: 
\[
\bvec\cdot\nabla\left(\frac{\Jpar}{B_0}\right) = \left(\bvec_0\cdot\nabla - \frac{1}{B_0}\bvec_0\cdot\nabla\apar\times\nabla\right) \left(\frac{\Jpar}{B_0}\right)
\]
Previously, time-evolution codes solving resistive and/or extended MHD have
been used to simulate ELMs \cite{sovinec-2004,brennan-2006,pankin-2007,snyder-2005,snyder-2006,huysmans-2007,park-1999,park-2007}.
To our knowledge, these are the first ideal MHD time-dependent simulations of ELMs: no dissipation
is intentionally introduced, so the only dissipation present is numerical.

Boundary conditions used for the simulations presented here are:
\[
\omega = 0 \qquad \delp\apar = 0 \qquad \deriv{P}{\psi} = 0 \qquad \phi = 0
\]
on inner and outer radial boundaries.

The coordinate system used for these simulations is that given in section~\ref{sec:tok_coord}, 
a field-aligned (flux) coordinate system with shifted radial derivatives. Differencing methods
used are $4^{th}$-order central differencing and $3^{rd}$-order WENO advection scheme. Radial boundary
conditions used are zero-gradient pressure perturbation, zero parallel current, and zero vorticity.
For these simulations no X-point is included and so the domain
is periodic in $y$ (with a twist-shift condition, see section~\ref{sec:tok_coord}) and periodic in $z$ (toroidal
angle). For efficiency, when performing linear simulations of a single
toroidal mode number $n$, only $1/n^{th}$ of the torus is simulated. Non-linear
effects will couple toroidal mode-numbers, and so a larger fraction of the torus
must then be simulated.

\subsection{Linear benchmarking}
\label{sec:elm_lin}

In order to benchmark BOUT++ for this problem, linear simulations have been
performed and comparison made to the ELITE linear code \cite{snyder-2002,wilson-2002}.
The data shown here is from a grid with 256 radial points ($x$), 64 poloidal ($y$) and
16 toroidal ($z$).
One difficulty in comparing linear MHD codes with time-evolving simulations is the treatment of the
vacuum region surrounding the plasma: whereas linear codes can treat this region analytically using
Green's functions, time-dependent codes must simulate quantities in this region and handle a
moving plasma-vacuum interface. For the results presented here, no 
distinction is made between the plasma and vacuum regions, and the test case used is strongly
ballooning (pressure-driven) rather than peeling (current-driven). This makes the solution relatively
insensitive to the location of any plasma-vacuum interface and provides a simplified starting point
for comparison. Inclusion of a vacuum region, possibly following the level-set methodology used in
NIMROD \cite{kruger-2001}, and simulation of the peeling instability is the subject of future work.

\begin{figure}[htb!]
\centering
\includegraphics[scale=0.4]{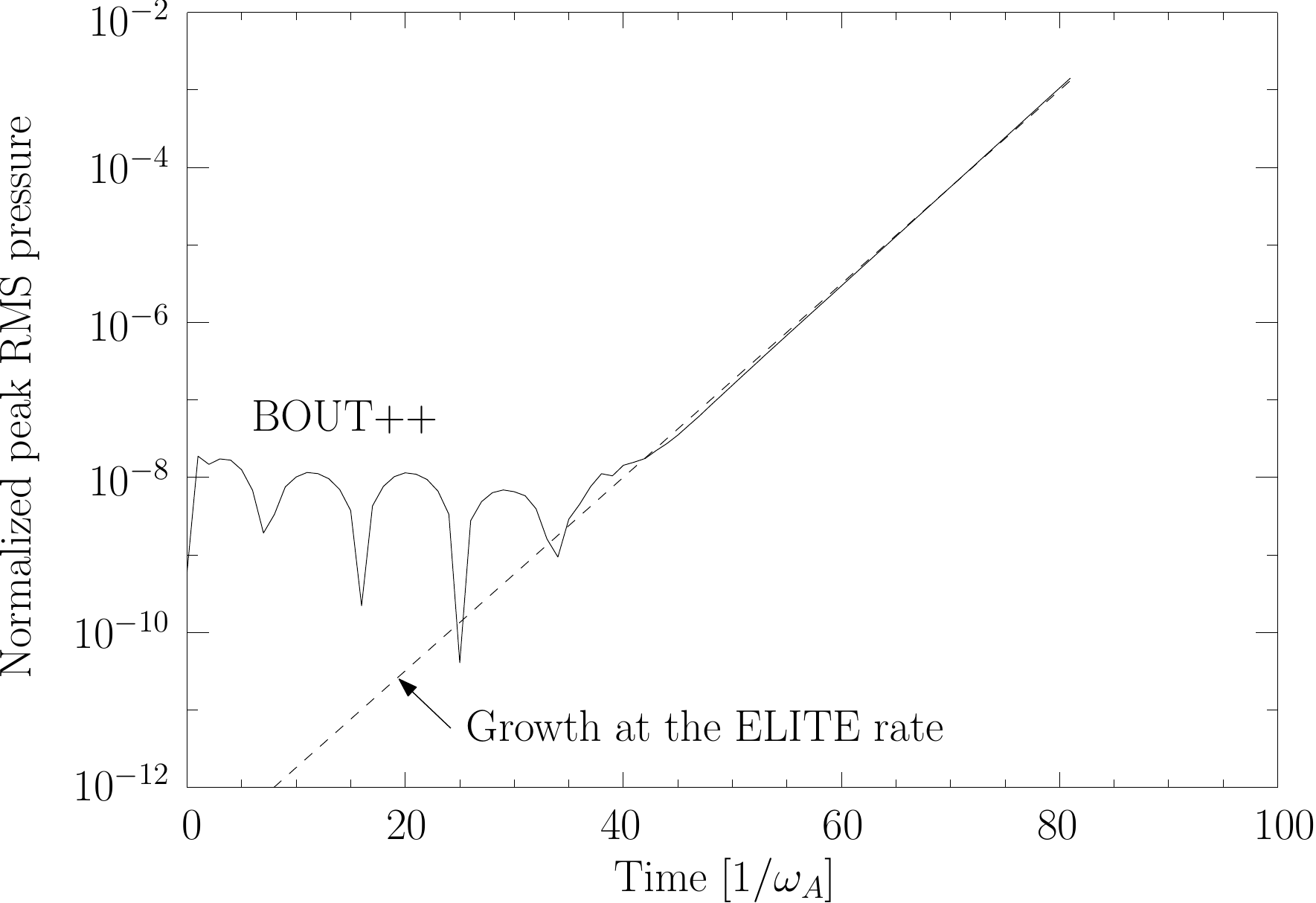}
\caption{Peak RMS pressure perturbation in BOUT++ giving a growth rate $\gamma = 0.245/\omega_A$. Comparison to ELITE growth-rate of $\gamma = 0.239/\omega_A$ (dashed line)}
\label{fig:n20_rms_p}
\end{figure}
The BOUT++ initial perturbation has a single toroidal mode-number $n=20$, but
is not an eigenmode of the system, and so first evolves through a transient
phase before settling on an eigenmode with a single
growth-rate. This can be seen in figure~\ref{fig:n20_rms_p} which shows the time-evolution
of the peak RMS pressure. The growth-rate the solution settles on is $\gamma = 0.245\omega_A$,
close to the ELITE result of $0.239\omega_A$ (shown as a dashed line in
figure~\ref{fig:n20_rms_p}). These growth-rates (and the time axis of
figure~\ref{fig:n20_rms_p}) are normalised to an Alfv\'en frequency
$\omega_A = V_A / R$, with $V_A=B_0 / \sqrt{\mu_0\rho_0}$, and $R$ the major radius.

There are several differences between BOUT++ and ELITE
which could explain the small growth-rate discrepancy, including the
equations solved, and the handling of the vacuum region:
both BOUT++ and ELITE solve a reduced
form of MHD valid for high-n, and should be identical in the limit of toroidal
mode number $n\rightarrow\infty$, but ELITE uses the energy principle
\cite{hazeltine-2003} rather than time-evolution, and so is reduced from MHD
in a different way to BOUT++. Differences are therefore expected for finite $n$.
A major difference is in the handling of the vaccum region: whereas ELITE uses an analytic calculation for the
vacuum contribution, currently BOUT++ treats the ``vacuum'' region as a
low-pressure ideal plasma. Future work includes improving modelling of this
vacuum region, which is essential for the correct simulation of peeling modes.

Linear MHD codes such as ELITE calculate the mode structure of an instability in terms of the
displacement vector $\mathbf{\xi}$, so that the plasma velocity is given by $\gamma\mathbf{\xi}$. 
The radial component $\xi_\psi$ is related to the electrostatic potential $\phi$ calculated by BOUT++
through the $\mathbf{E}\times\mathbf{B}$ velocity:
$\gamma\xi_\psi = -\nabla\phi\times\mathbf{B}/B^2 \cdot \mathbf{\hat{e}}_\psi$. 
The conversion from $\phi$ to $\xi$ is simplified because in ideal MHD the
frequency of an unstable mode is entirely imaginary - there is no real frequency component.
If diamagnetic or other non-ideal effects are included, then a real frequency
component appears and must be taken into account.

\begin{figure}[htb!]
\centering
\subfigure[BOUT++]{
  \label{fig:pol_n20_bout}
  \includegraphics[scale=0.6]{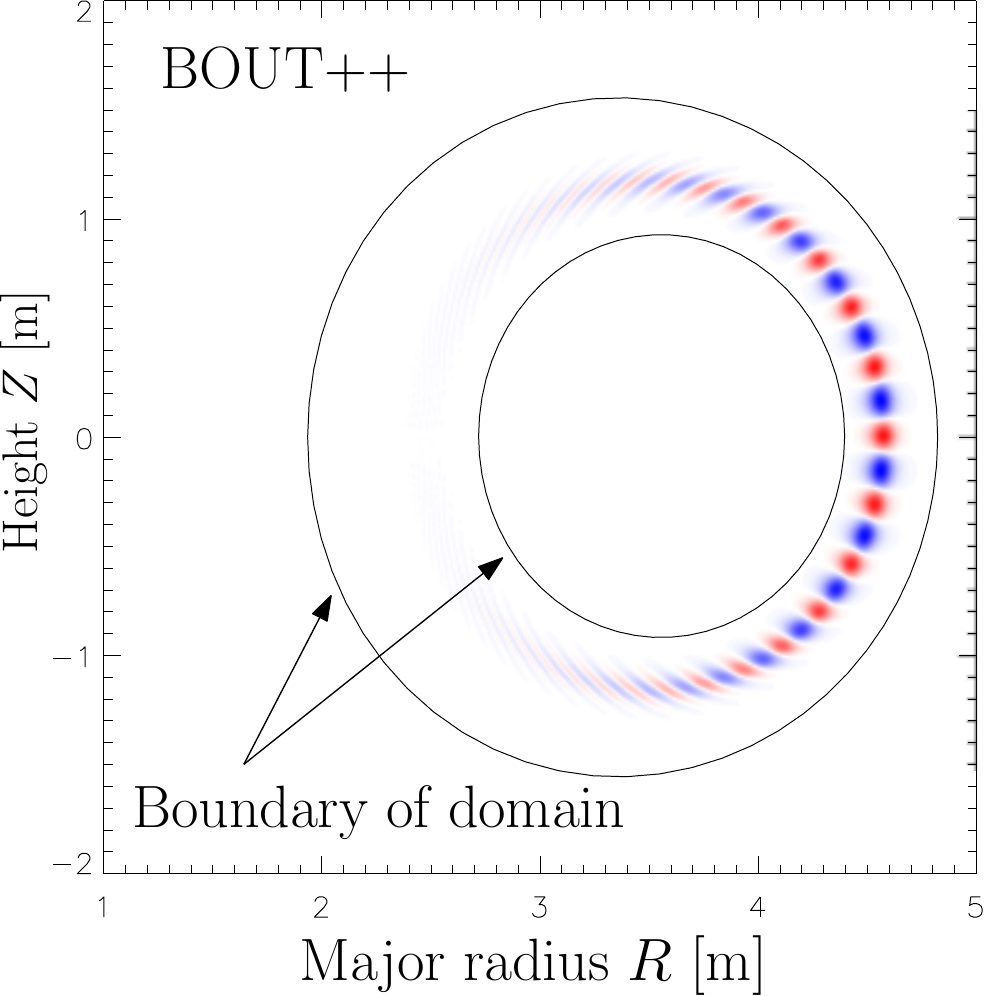}
}
\subfigure[ELITE]{
  \label{fig:pol_n20_elite}
  \includegraphics[scale=0.6]{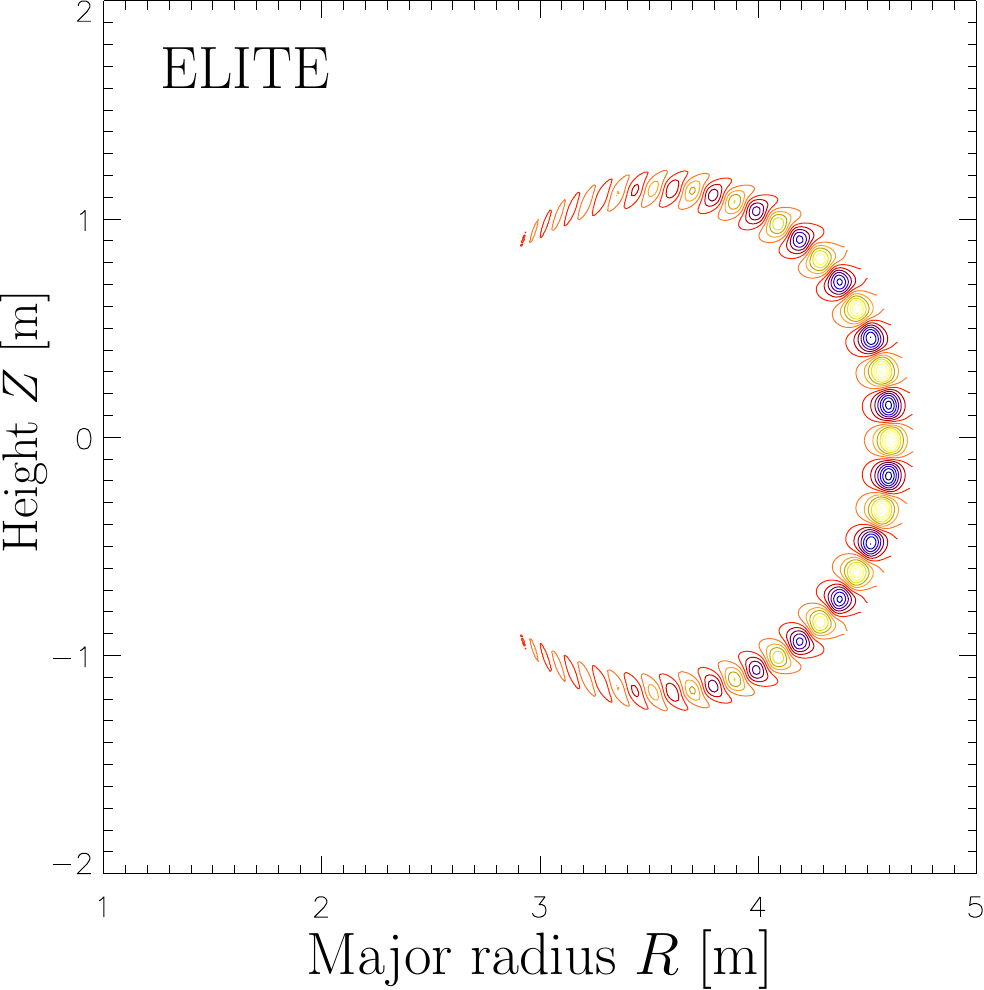}
}
\caption{Poloidal slice through the tokamak, showing radial displacement $\xi_\psi$ for toroidal mode number $n=20$.}
\label{fig:pol_n20}
\end{figure}
Figure~\ref{fig:pol_n20} shows the radial displacement $\xi_\psi$ on a
poloidal slice ($\psi, \theta$) through the tokamak at fixed toroidal
angle $\zeta$ from both BOUT++ (left) and ELITE (right). 
Note that due to the field-aligned coordinate system, the BOUT++ $\theta$
resolution at fixed $\zeta$ shown in figure~\ref{fig:pol_n20} is much higher
than the resolution in $y$: the number of $y$ poloidal points used (64) is
actually the number of points along a given field-line, rather than the resolution in $\theta$. Instead,
the $\theta$ resolution of figure~\ref{fig:pol_n20} is determined by the number of field-lines simulated ($n_z = 16$),
the number of times these are repeated to form a torus ($n = 20$), and the number of times each field-line travels around
the torus toroidally for each poloidal revolution (safety factor $q$). In this
case, $1.29 < q < 5.39$ (depending on radial location), and so the $\theta$ resolution is
$n_\theta = q n n_z \simeq 414 \rightarrow 1726$. 
This illustrates the advantages of using field-aligned coordinate systems.

A commonly used, and more quantitative way to display the information in figure~\ref{fig:pol_n20}
is in terms of the poloidal mode structure. This is the amplitude of poloidal Fourier modes,
calculated by taking the Fourier transform in a poloidal angle $\chi$:
\[
\chi = \frac{1}{q}\int_{\theta_0}^{\theta}\frac{\Bt}{\Bp R} \sqrt{\left(\deriv{R}{\theta}\right)^2 + \left(\deriv{Z}{\theta}\right)^2} d\theta
\]

\begin{figure}[htb!]
\centering
\subfigure[BOUT++ $\xi_\psi$ mode structure]{
  \label{fig:mode_n20_bout}
  \includegraphics[scale=0.6]{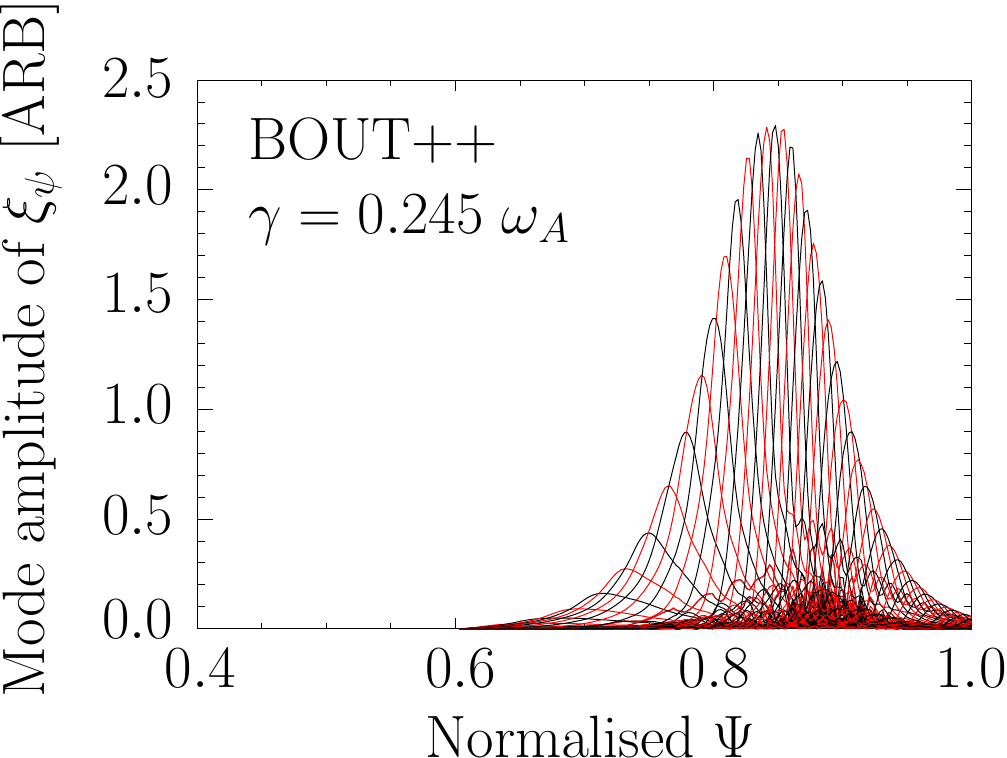}
}
\subfigure[ELITE $\xi_\psi$ mode structure]{
  \label{fig:mode_n20_elite}
  \includegraphics[scale=0.6]{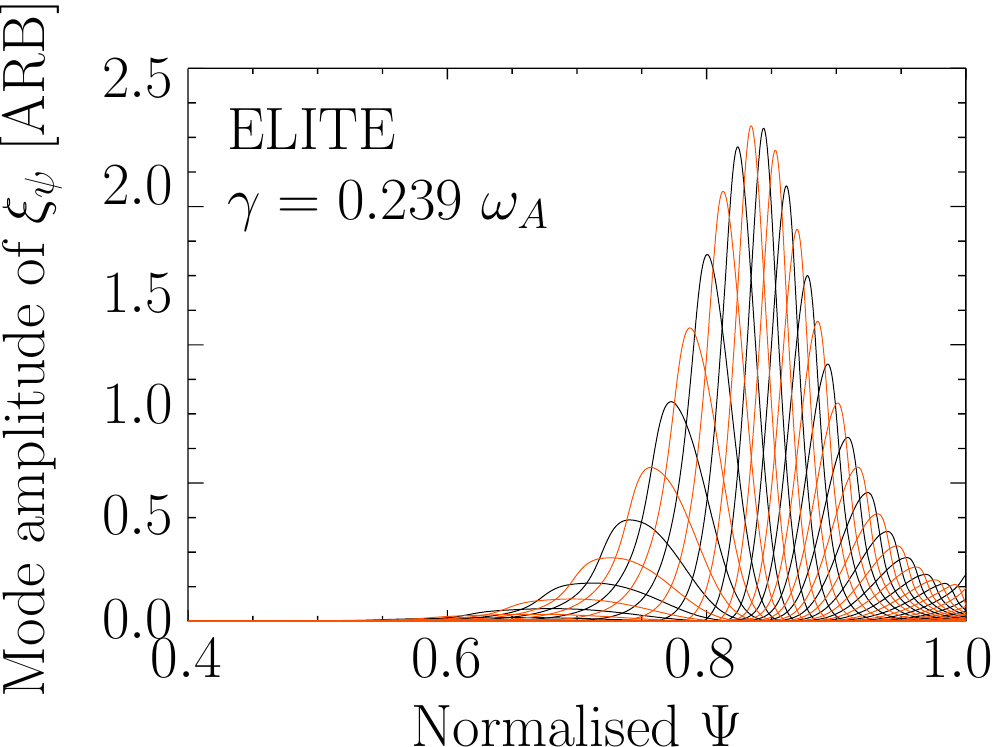}
}
\caption{Mode structure for toroidal mode number $n=20$}
\label{fig:mode_n20}
\end{figure}
Figure~\ref{fig:mode_n20} shows the mode-structure calculated by BOUT++ and ELITE for a test case with a 
toroidal mode number of $n=20$. Each line in this figure represents the magnitude of a poloidal
harmonic against the normalised poloidal flux $\psi$ ($\psi =0$ at magnetic axis, $1$ at plasma edge).
The BOUT++ domain covers the range $0.6 < \psi < 1.2$, but is shown in figure~\ref{fig:mode_n20}
on the same scale as ELITE for comparison.

\begin{figure}[htb!]
\centering
\includegraphics[scale=0.4]{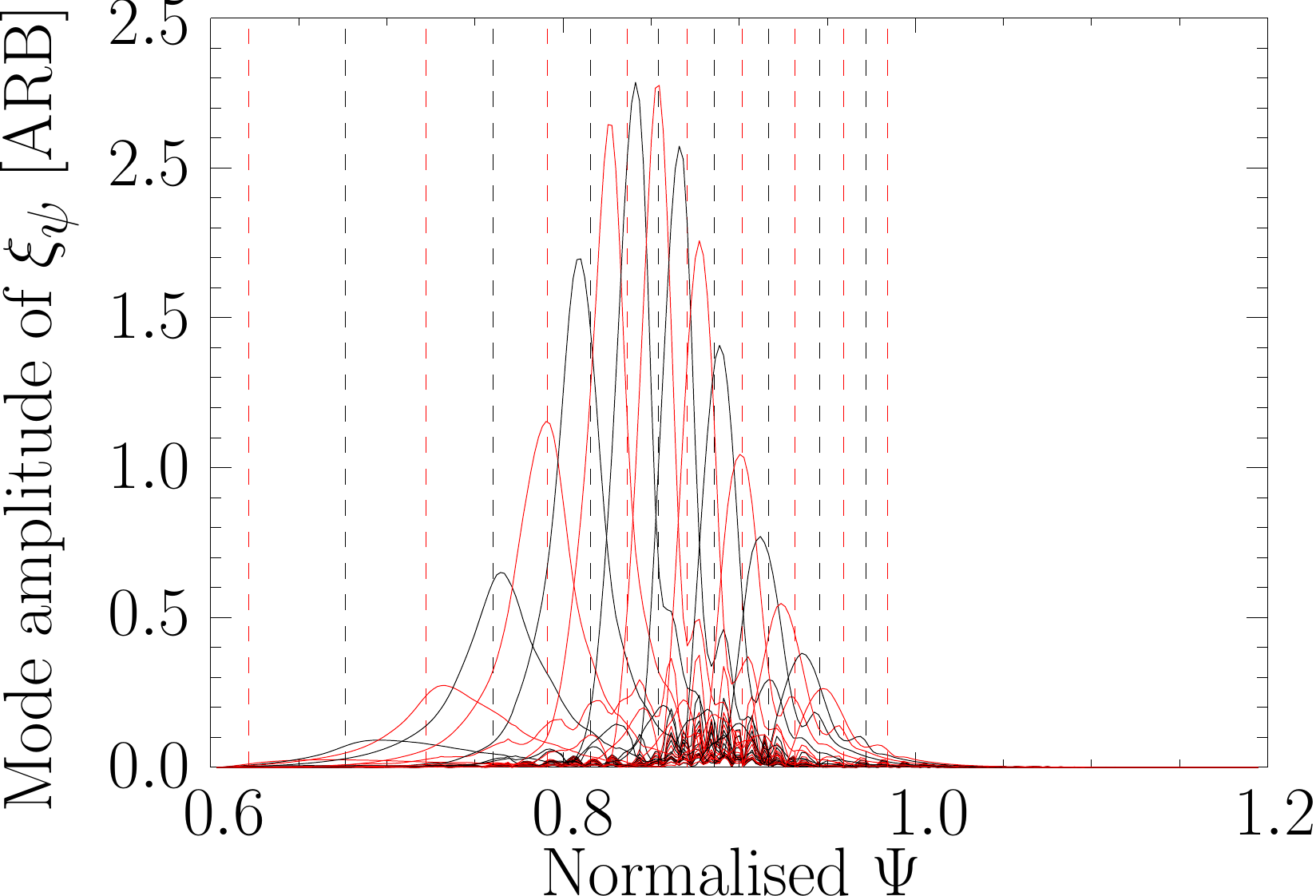}
\caption{Subset of poloidal modes in figure~\ref{fig:mode_n20} from BOUT++ simulation.
Vertical lines mark the location of resonant magnetic surfaces}
\label{fig:subset_phi}
\end{figure}
In both of these results the amplitude of a given poloidal mode peaks close to its resonant
magnetic surface, as is expected from analytic theory. This is shown in
figure~\ref{fig:subset_phi} which shows a subset of the BOUT++ poloidal modes
(every 2nd mode for clarity), with their resonant surfaces 
i.e. where $q\left(\psi\right)$ is rational. 

As expected for a time-dependent code, the BOUT++ result in figure~\ref{fig:mode_n20}
is less ``clean'' than the ELITE result, containing additional poloidal modes from the initial transient.
These additional modes gradually reduce in amplitude relative to the main resonant modes.
The mode envelope of the BOUT++ result in figure~\ref{fig:mode_n20} is in good agreement
with the ELITE result, but individual poloidal harmonics are slightly more peaked in the BOUT++ result.
The main difference is close to the plasma ``boundary'' at normalised $\psi = 1$. This is likely to be because
ELITE is treating the region beyond this point as a vacuum, whereas BOUT++ treats it as an ideal plasma.
Future work will include using different models of this vacuum region to assess its impact.

Linear time-dependent simulations using BOUT++ currently reproduce many of the features of
peeling-ballooning modes expected from analytic theory. The growth-rate and mode-structure
produced in the BOUT++ simulation is very close to that from the linear MHD code ELITE.
This is a proof-of-principle which demonstrates that BOUT++ is capable of simulating
the ideal ballooning mode correctly using reduced ideal MHD.
Further benchmarking against ELITE for other mode-numbers and equilibria, and non-linear
ELM simulations are the subject of a future paper.

\section{Performance}
\label{sec:perform}

Although the priority for BOUT++ is flexibility, it is also aimed at performing large-scale
($10^7 - 10^8$ variable) simulations. Therefore, speed of execution and scaling to large number
of processors is also important (see discussion in section~\ref{sec:opt}).
Here scaling of BOUT++ run-times with problem size and number of processors (hard scaling)
are described. Except where otherwise stated, the linear ELM problem is used as the test case.

\subsection{Scaling with problem size}

Typically 80-90\% of the wall-clock time is spent evaluating the time-derivatives calculated
in the physics module, with the remainder of the time spent advancing the time in the implicit CVODE
solver. Because the time-integration scheme used is implicit, the time-step is not limited by
the fastest waves on the grid. Instead, the time-step is determined by the accuracy
and hence behaviour of the simulation. This makes scaling with problem size harder
to quantify than an explicit code where the CFL condition determines the time-step. 
This test is therefore problem-dependent, and is intended as a guide to the performance
of BOUT++.

\begin{figure}[htbp!]
\centering
\subfigure[Scaling with $x$ dimension. $n_y=64$, $n_z=16$]{
  \label{fig:scale_x}
  \includegraphics[scale=0.35]{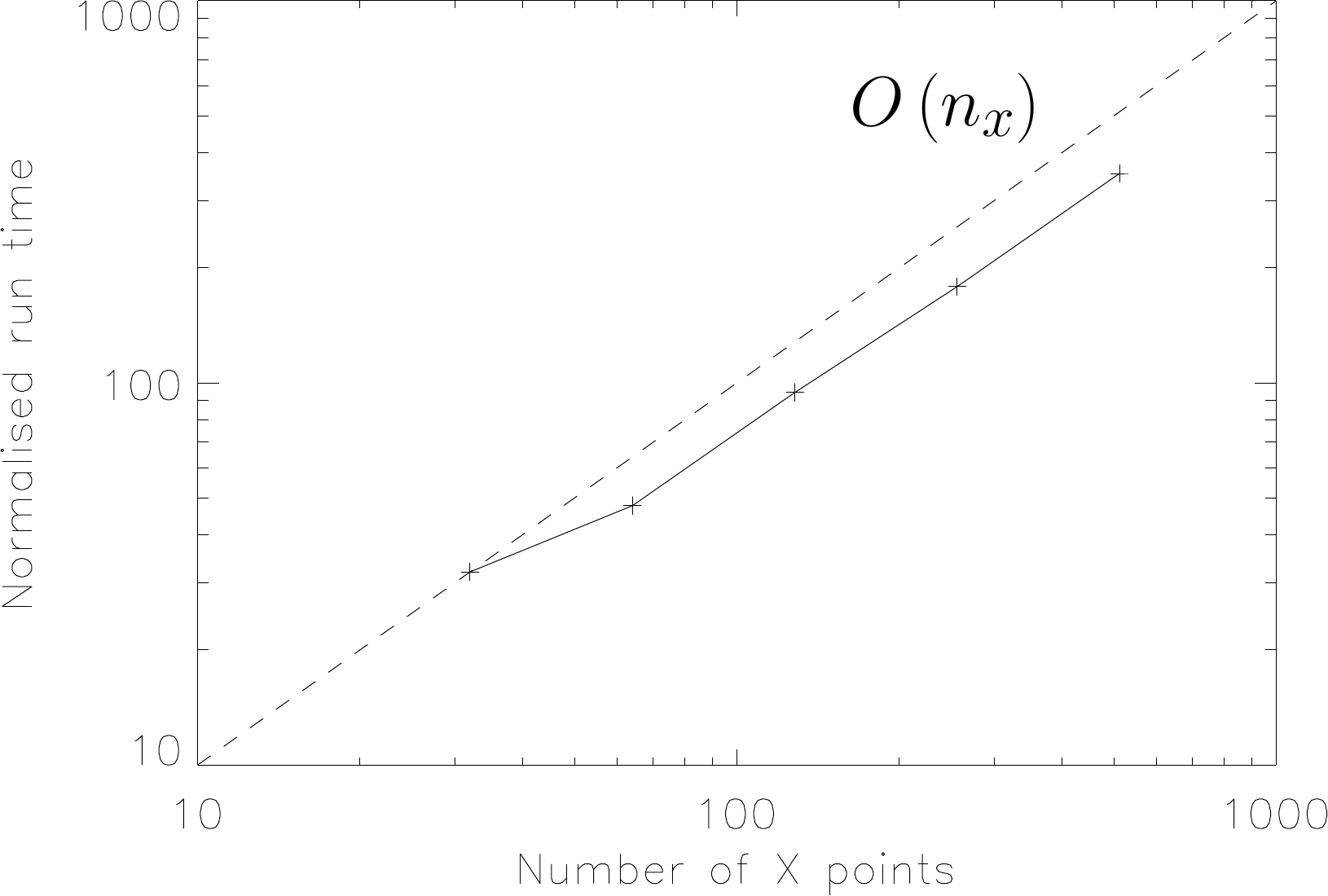}
}
\subfigure[Scaling with $y$ dimension. $n_x=128$, $n_z=16$]{
  \label{fig:scale_y}
  \includegraphics[scale=0.35]{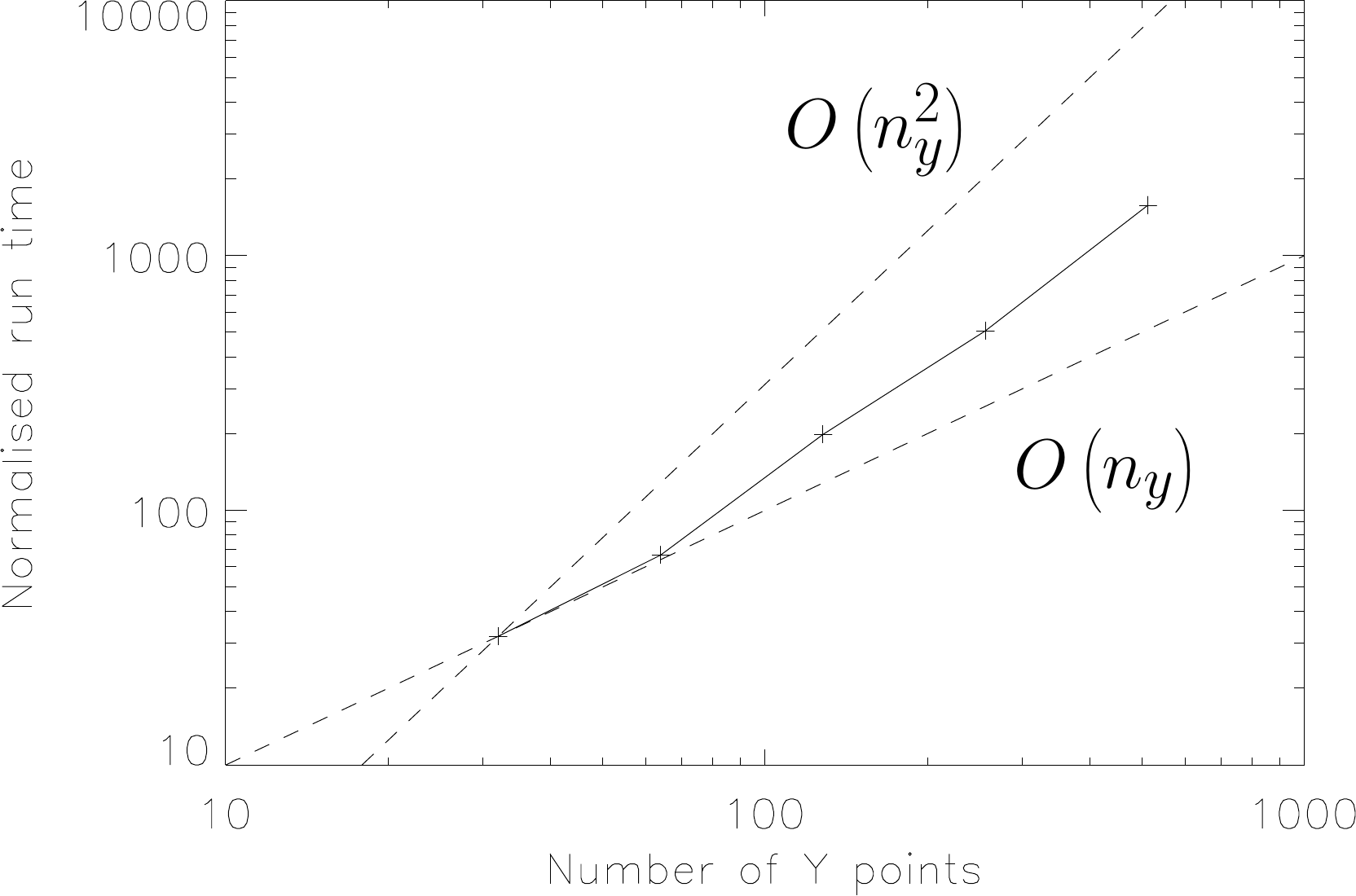}
}
\subfigure[Scaling with $z$ dimension, showing linear dependency. $n_x=256$, $n_y=64$]{
  \label{fig:scale_z}
  \includegraphics[scale=0.35]{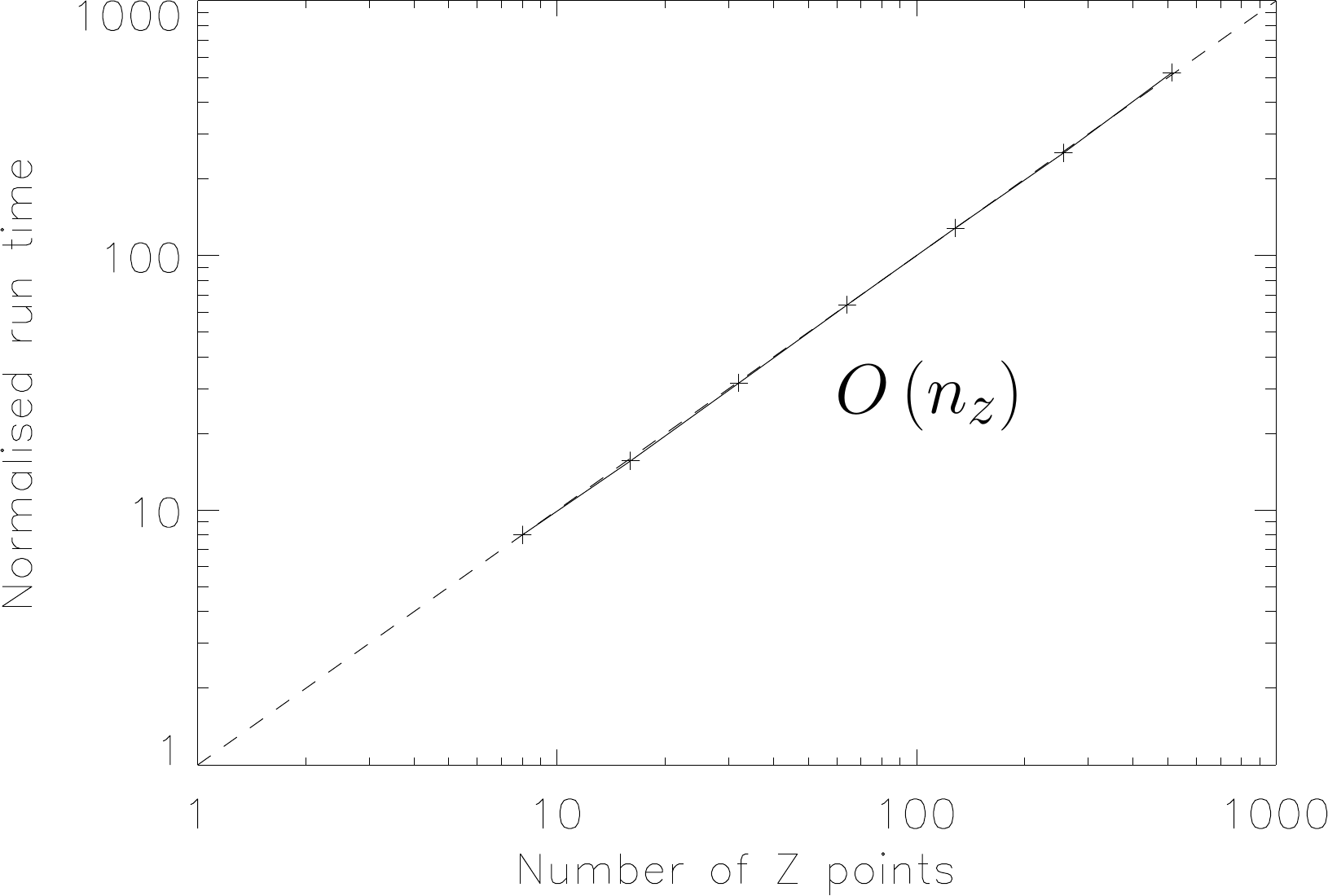}
}
\caption{Run time scaling with problem size on fixed number of processors (4) }
\label{fig:scaling_size}
\end{figure}

Figure~\ref{fig:scaling_size} shows variation of run-times on a fixed number of processors
with problem size in the $x$, $y$ and $z$ directions on a log-log scale.
The test case used is the linear ELM simulation described
in section~\ref{sec:elm_lin}. Scaling with $x$ and $z$ domain size is approximately
linear (dotted line) since in this case the dynamics in these directions are not
limiting the time-step. Scaling in the $y$ direction
(figure~\ref{fig:scale_y}) is between $O\left(n\right)$ and $O\left(n^2\right)$,
since fast parallel dynamics do have an impact on the time-step.
These tests show that the algorithms used are efficient ($O\left(n\right)$) where
the grid-size does not affect the time-step, and that at worst the scaling with
problem size is $O\left(n^2\right)$.

\subsection{Scaling with number of processors}
\label{sec:scale_proc}

Scaling of the BOUT++ code over a varying number of processors and fixed problem size
(hard scaling) has been performed
using the National Energy Research Scientific Computing (NERSC) Franklin
machine. This is a Cray XT-4 with 9,660 nodes linked by SeaStar 2 interconnect.
Each node consists of a dual-core 2.6GHz AMD Opteron, giving a theoretical peak
performance for Franklin of approximately 101.5 TFlop/s.

Scaling efficiency $\epsilon$ on $N_P$ processors is given relative to a reference number of processors $N_{P0}$
using the run-time $T\left(N_P\right)$ as:
\[
\epsilon\left(N_P\right) = 100 \frac{N_{P0} T\left(N_{P0}\right)}{N_P T\left(N_P\right)}
\]
which gives an estimate of the percentage of CPU time used in solving the problem, rather than
syncronising with other processors.

\begin{figure}[htbp!]
\centering
\subfigure[Efficiency]{
  \label{fig:scaling2_eff}
  \includegraphics[scale=0.4]{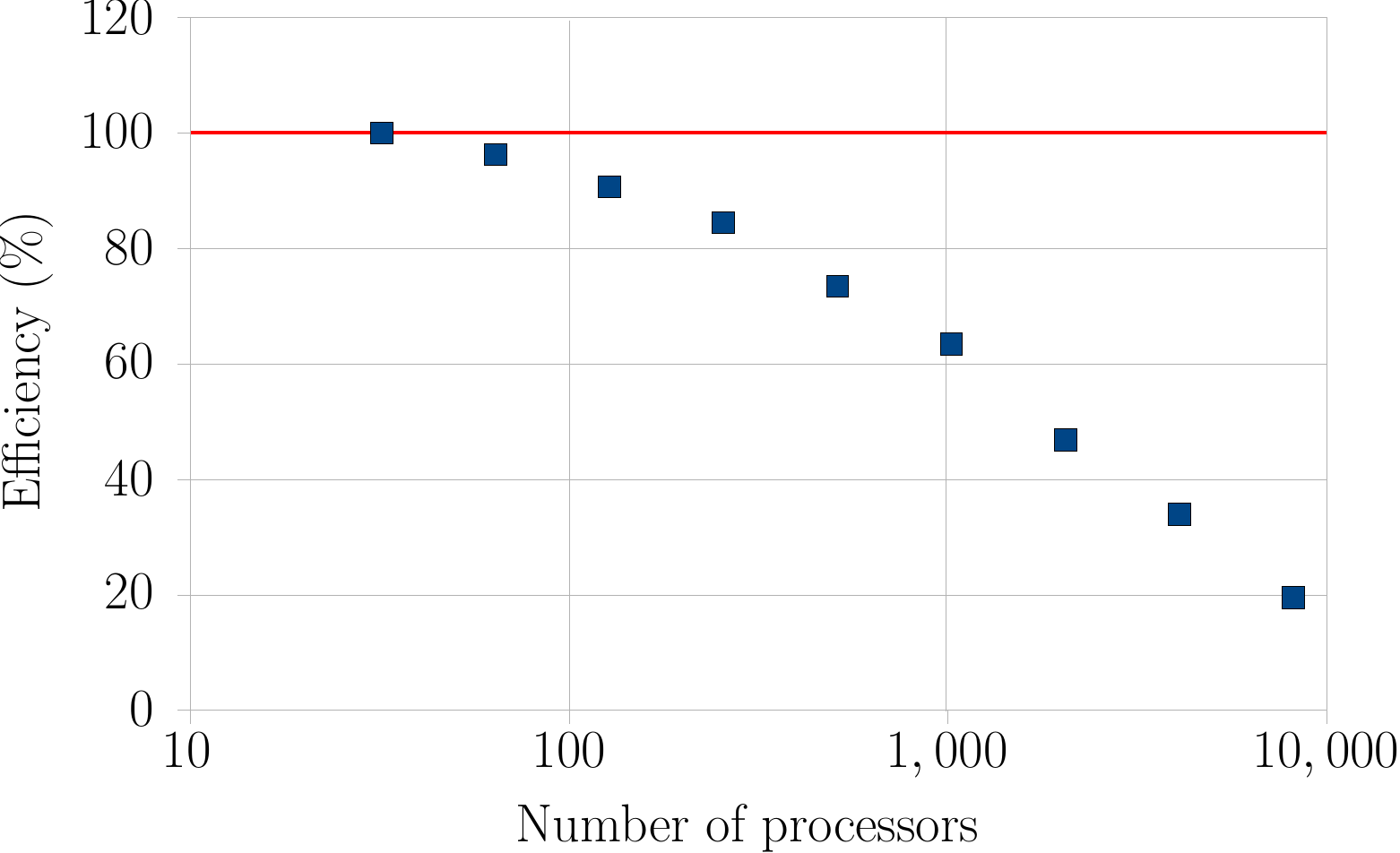}
}
\subfigure[CPU usage]{
  \label{fig:scaling2_use}
  \includegraphics[scale=0.5]{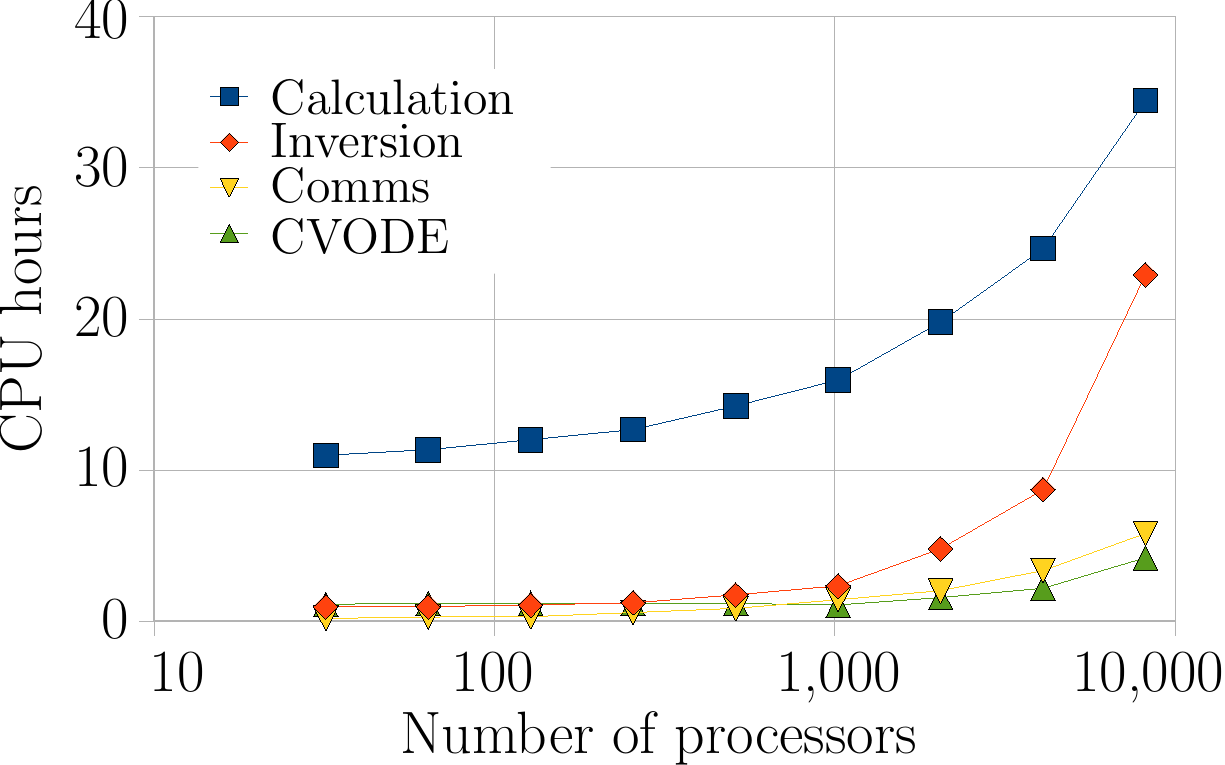}
}
\caption{Scaling of a 3D ELM simulation on a $256\times 256\times 128$ mesh. CPU usage is given for arithmetic and differencing operations(blue); Laplacian inversion code (red); communications not part of inversion (yellow) and CVODE the time-advance code (green)}
\label{fig:scaling2}
\end{figure}

Figure~\ref{fig:scaling2} shows the scaling of an ELM simulation
solving 3 fields on a 256x256x128 grid (i.e 25,165,824 evolving variables) on up
to 8192 processors. This shows that for this problem 2048 processors can be 
used at approximately 50\% efficiency relative to 32 processors. The number of
processors which can be efficiently employed is dependent on the problem size,
so this is the approximate grid size anticipated for future non-linear ELM simulations.
For the linear ELM simulations in section~\ref{sec:elm_lin} ($256\times 64\times 16$ mesh),
the calculation is 50\% efficient for between 256 and 512 processors.

One of the bottlenecks in ELM simulations for large numbers of processors
is the Laplacian inversion code: figure~\ref{fig:scaling2_use} shows the number of
CPU hours devoted to various parts of the code, showing that the time spent in this inversion code
(red, diamonds) becomes significant for large number of processors. Currently the Laplacian inversion
algorithm is quite a simple
adaptation of the Thomas serial tridiagonal inversion scheme. The calculation is
parallelised by performing inversion of Y slices simultaneously,
but efficiency will decline once the number of processors in X
exceeds the number of poloidal (Y) points per processor:
$P_X > N_Y / P_Y$ i.e. $P > N_Y$. This may explain the faster decline
in efficiency for greater than 256 processors in figure~\ref{fig:scaling2_eff}.

This Laplacian inversion code was implemented because it performs exactly
the same operations as the serial code, and so provides a good base case
for testing. The Thomas algorithm is however notoriously inefficient on
parallel computers and several better algorithms exist and
will be implemented in future. 

In order to test the efficiency of the code in the absence of Laplacian inversions, 
a scaling study has been done for the 2D Orszag-Tang vortex problem in section~\ref{sec:otv}:
Full ideal MHD on a $512\times 512$ mesh with 2,097,152 evolving variables.
\begin{figure}[htbp!]
\centering
\subfigure[Efficiency]{
  \label{fig:scaling3_eff}
  \includegraphics[scale=0.4]{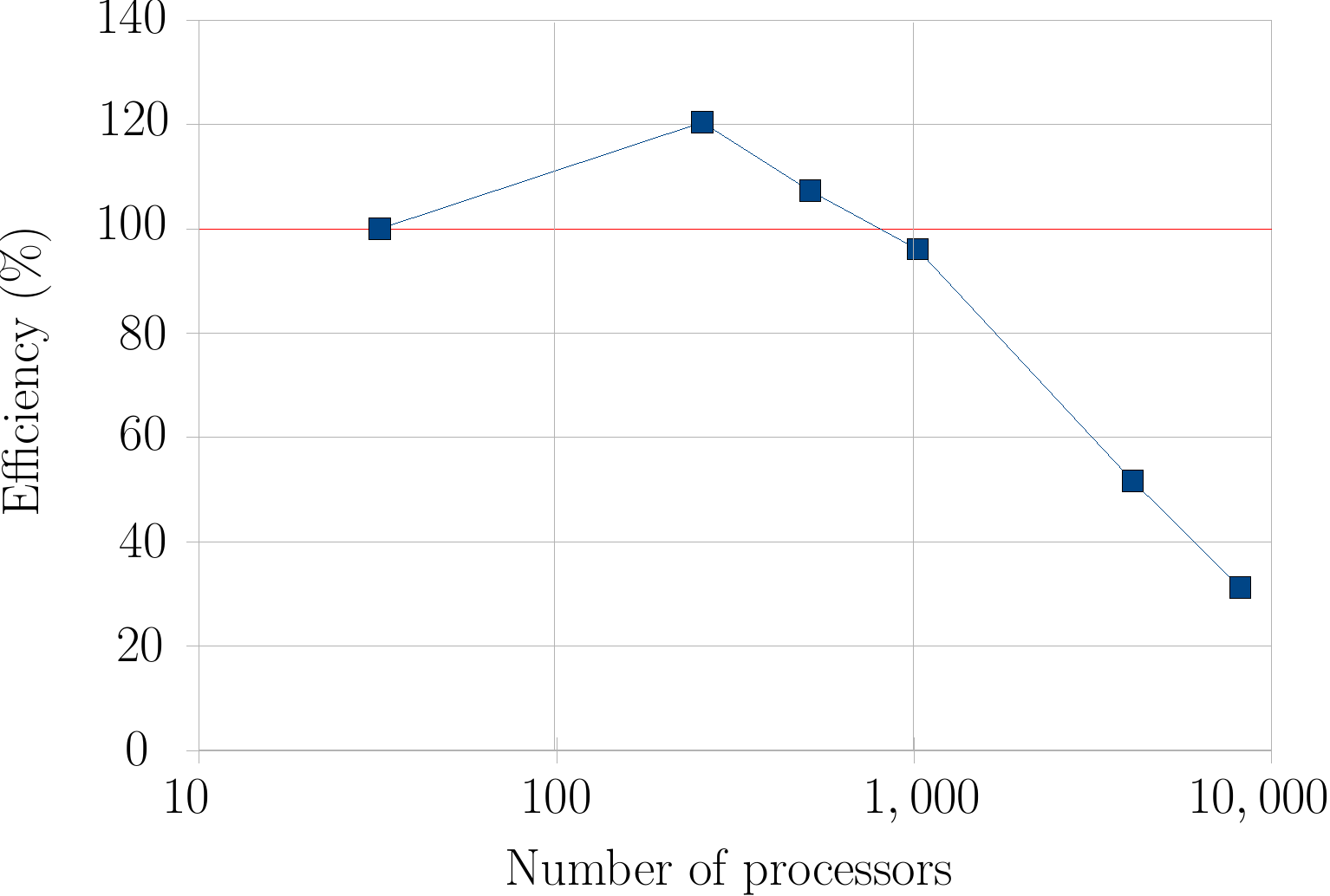}
}
\subfigure[CPU usage]{
  \label{fig:scaling3_use}
  \includegraphics[scale=0.5]{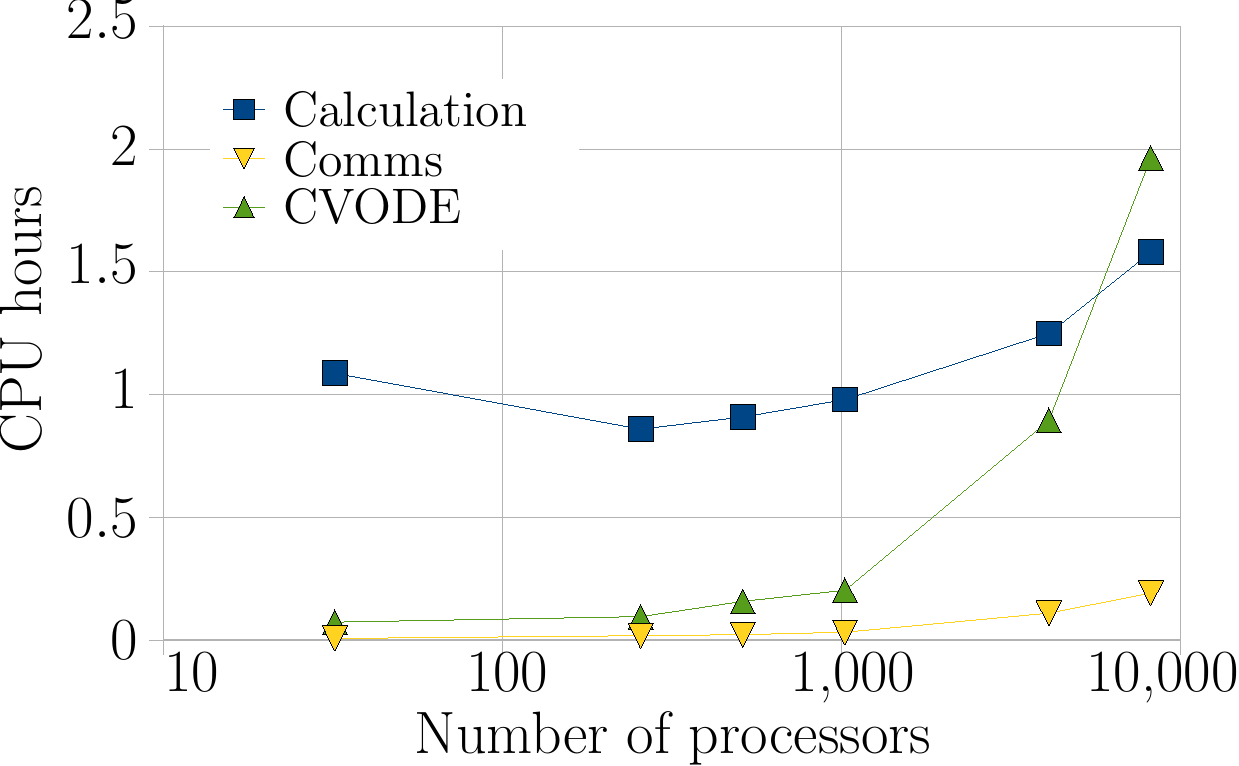}
}
\caption{Scaling of 2D ideal MHD Orszag-Tang vortex problem on a $512\times 512$ mesh}
\label{fig:scaling3}
\end{figure}
Figure~\ref{fig:scaling3_eff} shows the efficiency of this case (again relative to 32 processors).
In this case - despite having fewer evolving variables - scaling is over 50\% efficient
for 4096 processors.
Note the increase in efficiency between 32 and 1024. This may be due to the increasing
amount of available cache: smaller number of grid-points per processor mean more data
will fit into fast cache, reducing memory access times. This may indicate that memory
access is taking up a significant part of the processing time, which would be expected to be more of a problem
for BOUT++ than its more specialised predecessor BOUT since BOUT++ performs a loop over grid-points for every
operation, whereas BOUT loops over grid-points once.

For this case the total number of CPU hours used for each task is shown in figure~\ref{fig:scaling3_use}.
Communication of variables (yellow) doesn't seem to be a significant problem for any 
of these test cases; in this case the limiting factor appears to be CVODE (green).
This may be related to the small number of variables; it's probable that CVODE could
use more processors efficiently given a larger problem to solve. 

\section{Conclusions}

A new fluid simulation code BOUT++ has been developed and some tests presented.
The code is very modular, allowing new features such as differencing methods
to be quickly implemented. In particular the fluid model solved can be easily
changed to include an arbitrary number of scalar and vector fields. 
This allows BOUT++ to be used as a platform for quickly testing both new algorithms,
and different physics models.

Numerical methods currently included are a
fully implicit solver (the CVODE library, section~\ref{sec:time_int}), 
and WENO schemes for handling shocks. The stability and accuracy of these schemes has been
demonstrated using a series of linear and non-linear problems (section~\ref{sec:tests}).
Whilst the current implementation has been found to be stable in the presence of shocks,
accuracy in the vicinity of shocks needs further improvement.

Increased flexibility often comes with a performance cost and so several
optimisation strategies used in BOUT++ to reduce this penalty whilst
retaining flexibility have been described in section~\ref{sec:opt}.
Scaling of run-times with problem size and processor number in
section~\ref{sec:perform} show the efficiency of the algorithms ($O\left(n\right)$
where the time-step is not impacted, worst case $O\left(n^2\right)$). Hard scaling
to thousands of processors has been demonstrated using NERSC's Franklin Cray XT4 machine.
Areas for improvement have been highlighted, particularly
the need for a faster parallel Laplacian inversion algorithm.

The main motivation in developing this code is to simulate Edge Localised Modes
(ELMs) in tokamaks, and so several features specific to tokamak geometry have been
implemented such as shifted radial derivatives described in
section~\ref{sec:tok_coord} and the topology necessary for simulation of equilibria
with x-points. Linear simulations of ELMs
reproduce many features of the ELM such as mode-structure and growth-rate very close
to that produced by the ELITE linear MHD code. These promising results indicate
that BOUT++ is capable of accurately simulating ELMs using reduced ideal MHD, the
first time such dissipationless (apart from numerical) simulations have been done. 
An extended analysis of BOUT++ ELM simulations, comparison with ELITE, and non-linear
behaviour is the subject of a future publication.
\bibliography{bout++}
\bibliographystyle{cpc}

\end{document}